\documentclass{article}

\usepackage{multirow}
\usepackage{PRIMEarxiv}
\usepackage{float}
\usepackage{authblk}
\usepackage{amsmath}
\usepackage{cancel}
\usepackage{makecell}
\usepackage{algorithm}
\usepackage{algpseudocode}
\usepackage{algpseudocode}
\usepackage[dvipsnames]{xcolor}
\usepackage[utf8]{inputenc} 
\usepackage[T1]{fontenc}    
\usepackage{hyperref}       
\usepackage{url}            
\usepackage{booktabs}       
\usepackage{amsfonts}       
\usepackage{nicefrac}       
\usepackage{microtype}      
\usepackage{lipsum}
\usepackage{fancyhdr}       
\usepackage{graphicx}       
\usepackage{caption}
\usepackage{subcaption}
\usepackage[titletoc]{appendix}
\usepackage{titlesec}
\usepackage{caption} 
\graphicspath{{media/}}     

\newcommand{\argmin}{\mathop{\arg\min}}

\newcommand{\K} {K'} 
\newcommand{\n} {n} 
\newcommand{\B} {B} 
\newcommand{\rr} {r} 
\newcommand{\Rset} {\mathbb{R}} 

\newcommand{\x}{x}

\newcommand{\pfrac}[2]{\frac{\partial #1}{\partial #2}} 
\newcommand{\ppfrac}[3]{\frac{\partial^2 #1}{\partial #2\partial #3}}

\newcommand{\numineq}{n_\text{ineq}}
\newcommand{\numeq}{n_\text{eq}}

\newcommand{\numgrid}{n_\text{grid}}
\title{A flexible and differentiable coil proxy for stellarator equilibrium optimization}
\author[1,2,3, *]{Lanke Fu}
\author[1,2]{Dario Panici}
\author[4]{Elizabeth J. Paul}
\author[3]{Alan A. Kaptanoglu}
\author[2]{Amitava Bhattacharjee}
\affil[1]{Princeton Plasma Physics Laboratory, Princeton, NJ}
\affil[2]{Department of Astrophysical Sciences, Princeton University, Princeton, NJ}
\affil[3]{Courant Institute School of Mathematics, Computing, and Data Science, New York University, New York, NY}
\affil[4]{Department of Applied Physics and Applied Mathematics, Columbia University, New York, NY}
\affil[*]{Corresponding email: lanke\_fu@outlook.com}

\begin{document}
\maketitle

\begin{abstract}
Balancing plasma performance and coil cost is a significant challenge when designing a stellarator power plant. Most current stellarator designs are produced through two-stage optimization: stage-1 for the equilibrium and stage-2 for a coil design that reproduces its magnetic configuration. Because few proxies connect both stages, two-stage optimization can produce plasmas that have high-quality physical properties but overly complex coils. In recent years, single-stage optimization has increasingly been used to optimize the plasma and coils simultaneously in order to improve the plasma-coil balance. However, all existing single-stage tools are specialized for filament coils, cannot model coil systems containing permanent magnets (PM) or dipole arrays, and continue to be challenged by numerical problems. The quasi-single-stage (QSS) optimization finds a middle-ground by integrating a coil optimization subproblem into stage-1 optimization. We present a flexible, differentiable coil complexity proxy based on the newly developed QUADCOIL coil optimization code. QUADCOIL is fast and can target realistic coil metrics and constraints that are unavailable to codes with comparable speed. We demonstrate the effectiveness and flexibility of the QUADCOIL proxy by presenting two QSS optimization studies. The first study produces a permanent magnet solution for the MUSE stellarator with $29\%$ fewer magnets than previous solutions. The second study produces a coil solution for the ARIES-CS stellarator with $27\%$ reductions in both peak and root-mean-square force.
\end{abstract}


\section{Introduction}

Stellarators are attractive three-dimensional (3D) fusion devices that generate a rotational transform with external coils rather than with a plasma current. Unlike tokamaks, stellarators do not require steady-state current drive. Therefore, they are not susceptible to the current-driven instabilities that can cause  disruptions in tokamaks~\cite{intro_JET_disruption}. Experimental evidence also shows that stellarators can operate at pressures beyond their linear instability thresholds\cite{intro_pressure_limit}. Therefore, stellarator power plants are likely to have simpler control systems, lower power recirculation, higher triple product, and higher energy efficiency than tokamaks~\cite{intro_Menard_2011}. 

The design of the magnetic coil system is a critical part of the design of a stellarator power plant. Coil engineering cost is a primary factor in the cost of a stellarator~\cite{strykowsky_engineering_2009, intro_w7x_delay}. The accuracy of the coil magnetic field is directly related to plasma performance. Coil complexity also influences the choice of stellarator equilibrium. Recent studies have shown that the choice of equilibrium has an order-of-magnitude impact on coil complexity~\cite{lanke_fu_global,kappel_magnetic_2024}. In addition, each equilibrium can prefer a different type of coil topology. For example, LHD, a device with a simple helical coil set, is difficult to reproduce with modular coils~\cite{ lanke_fu_global,kappel_magnetic_2024}. These factors make coil-plasma co-design a field of active study in stellarator physics.

There are two main methods for designing a stellarator. Most existing stellarator configurations are produced using a two-stage method. Stage-1 produces an optimized equilibrium with favorable physical properties, and stage-2 designs a coil-set for the equilibrium. Both stages involve costly, high-dimensional, non-convex optimization problems. However, designing the equilibrium and coils separately is still more tractable than designing both at once. Because few coil complexity and force proxies connect the two stages, the two stage method can result in equilibria with good physical properties and unrealistically complex coils. 

Single-stage methods for stellarator optimization have become popular in recent years~\cite{intro_single_vmec_fixed3,intro_single_nae,Henneberg_single_2021}. These methods perform plasma and coil optimization simultaneously to find a balance between plasma performance and coil engineering costs. When combined with equilibrium solvers that allow the flux surfaces to break, the single-stage method can also help design plasmas that are robust to coil manufacturing errors~\cite{smiet}. 

While proven successful, many challenges remain with single-stage optimization. At the time of writing, most single-stage tools are specialized for filament coils. Permanent magnet (PM) or dipole array stellarators, such as the MUSE PM stellarator experiment~\cite{intro_muse}, replace complex filament coils with arrays of identical permanent magnets or dipole coils~\cite{intro_dipole2}. The start-up companies Thea Energy~\cite{kruger2025coil,swanson2025scoping} and Stellarex~\cite{zarnstorff2023stellarex} have proposed using large arrays of relatively small dipole coils. These devices promise high field accuracy at low engineering costs, but there is no published study on the single-stage optimization of PM or dipole arrays. Merging coil and plasma optimization also brings new numerical challenges. Single stage optimization often requires good initial states to converge. It requires either maintaining both coil and plasma degrees of freedom or performing challenging free-boundary equilibrium calculations~\cite{conlin2024high, VMEC_free_86, Henneberg_single_2021}. In addition, filament coil optimization is non-convex~\cite{zhu2017new, wechsung2022precise} and does not guaranty a unique coil set for each equilibrium. While this affords more design freedom, it also introduces new local minima to the single-stage landscape, since there can be distant regions in the parameter space that correspond to the same plasma supported by different coil sets. 

Recent works suggest that a coil complexity proxy based on a winding surface model may address the aforementioned challenges. The winding surface model treats an entire coil set as a continuous sheet current on a known winding surface~\cite{merkel1987solution, landreman2017improved, elder_three-dimensional_2024}. While not as realistic as filament, discrete dipole, or PM models, the winding surface model can be configured to study both types of stellarator coil systems. Recent works have shown that winding surface proxies can predict the results of filament optimization~\cite{lanke_fu_global} and improve the complexity of PM arrays~\cite{qss_Yu2024-sr}. In addition, coil optimization is convex in the winding surface model. This guarantees a unique coil set for each pair of plasma and winding surfaces and makes winding surface optimizations orders of magnitude faster than filamentary optimization. 

\subsection{Contributions of this work}
We will show in this paper that a "quasi-single-stage" (QSS) optimization (recently innovated by Yu et al.\cite{qss_Yu2024-sr}) that integrates a winding-surface-based proxy into stage-1 optimization can offer the benefits of single-stage optimization at low computational costs, ranging from $10$ minutes to $5$ hours on a personal computer. For this, we develop the QUADCOIL code~\cite{lanke_fu_global} into a differentiable, flexible coil complexity proxy. In particular, QUADCOIL extends QSS optimization with more complex coil objectives, end-to-end differentiability using JAX, and a winding surface generator for producing well-behaved surfaces that can move with the plasma surface during a QSS optimization.  We demonstrate the effectiveness of QUADCOIL QSS optimization in Sections~\ref{sec: results: PM} and~\ref{sec: results: force} by (1) performing QSS optimization, starting from a known equilibrium; (2) discarding the winding surface solution; (3) designing permanent magnets or filamentary coils from scratch for both equilibria; (4) comparing their performance. Coils of the QSS equilibria consistently achieve higher performance than those of the non-QSS equilibria. This confirms that QUADCOIL-based QSS optimization can indeed promote features in the plasma surface that lead to better coils. We believe that the results of QSS optimization can serve as initial states for single-stage optimization, leading to faster convergence and further improvement in coil-plasma balance.

Part of this work addresses the issue that winding surface models exhibit limited objective choices and no support for constraints. This severely limits the physical quantities available to a winding surface model, as well as its control over the topology of the surface current distribution. The latter, notably, means that one often needs to produce and evaluate multiple winding surface solutions to create a coil set that satisfies all engineering requirements. These drawbacks limit the usefulness of the model as a coil complexity proxy. QUADCOIL extends the winding surface model to support non-convex quadratic penalties and constraints with minimal loss in speed by maintaining "near-convexity". This makes it possible to use QUADCOIL as a flexible coil proxy that can be configured to model a range of realistic coil optimization problems, such as low-force filament coils and low-density dipole arrays.

The paper is organized as follows. Section~\ref{sec: theory} provides an introduction to the theory of stellarator optimization, winding surface models, and QUADCOIL. Section~\ref{sec: numerical} discusses the numerical methods for solving the QUADCOIL problem, differentiating its solutions, and robustly generating well-behaved winding surfaces. These methods enable QUADCOIL as a differentiable coil proxy. Section~\ref{sec: results} presents two QSS studies that validate the effectiveness of the QUADCOIL proxy. The first study presents two MUSE-like vacuum fields optimized for low permanent magnet count and density. The second study presents an ARIES-CS-like equilibrium optimized for low-force filament coils. Section~\ref{sec: conclusion} provides conclusions and possible avenues for future developments.

\section{Theory}\label{sec: theory}

\subsection{Stellarator optimization}
In this section, we will briefly introduce the formulation and characteristics of two-stage, single-stage, and quasi-single-stage optimization. We will also introduce the notation used in the rest of this paper.  

In this paper, primed quantities always represent variables associated with the coils. The coil parameters, which describe their geometry and currents, are denoted by $x'\in\mathbb{R}^N$. The equilibrium parameters, which often represent the shape of the plasma boundary, are denoted by $x\in\mathbb{R}^M$. The plasma objective and constraint functions are denoted by $f_p(x):\mathbb{R}^M\to \mathbb{R}$, $g_p(x):\mathbb{R}^M\to \mathbb{R}$, and $h_p(x):\mathbb{R}^M\to \mathbb{R}$. The coil objective and constraint functions are denoted by $f_c(x'):\mathbb{R}^N\to \mathbb{R}$, $g_c(x'):\mathbb{R}^N\to \mathbb{R}$, and $h_c(x'):\mathbb{R}^N\to \mathbb{R}$. Typically, $N\sim500$ degrees of freedom, while $M\sim100$. Although the precise numbers may vary, the coil parameters often contain more degrees of freedom than the equilibrium parameters~\cite{intro_single_vmec_fixed3}.

\subsubsection{Two-stage optimization}

The two-stage method obtains the coil and plasma solutions by solving two separate problems:
\begin{equation}\label{eq:two-stage}
    \begin{split}
        \text{Stage-1 (equilibrium): } x_* = \argmin_x\,&f_p(x),\\
        \text{subject to }&g_p(x)\leq0,\\
        &h_p(x)=0,\\
        \\
        \text{Stage-2 (coil): } x_*' = \argmin_{x'}\,&f_c(x', x^*),\\
        \text{subject to }&g_c(x')\leq0.\\
        &h_c(x')=0.\\
    \end{split}
\end{equation}
Here, the stage-1 problem often incorporates a fixed-boundary (or alternative) equilibrium solver as one of the constraints in $h_p$. Compared to single-stage methods, the main advantage of the two-stage method is its tractability. While both stages are still high-dimensional non-convex problems, separating the equilibrium and coil stages makes both problems simpler to solve. Nonetheless, many local and suboptimal minima exist, and recent evidence suggests that coil complexity can vary by orders of magnitude across combinations of equilibrium and coil topology~\cite{lanke_fu_global,kappel_magnetic_2024}. Because few coil proxies exist for the equilibrium stage, the two-stage method can produce equilibria that are difficult to support with simple coil sets.

\subsubsection{Single-stage optimization}
Single-stage stellarator optimization simultaneously optimizes the plasma and coil geometry to improve coil-plasma balance. Based on their formulations and equilibrium solvers, single-stage methods can be further divided into two categories:
\begin{enumerate}
    \item \textit{Fixed-boundary single-stage methods}: As the name suggests, these methods use fixed-boundary equilibrium solvers that require the plasma boundary as input. Because of this, a fixed-boundary single-stage method considers both the plasma boundary and coil geometries as degrees of freedom. In our notation, this is:
\begin{equation}\label{eq:fixed-boundary}
\begin{split}
   x_*, x'_* = \argmin_{x, x'}&\left[f_p(x) + f_c(x') + f_\text{match}(x, x')\right],\\
        \text{subject to }\quad &g_p(x)\leq0,\quad
        g_c(x')\leq0.\\
         &h_p(x)=0,\quad
        h_c(x')=0.\\
\end{split}
\end{equation}
Here, $f_\text{match}$ is a matching term ensuring that the plasma boundary is consistent with the magnetic field generated by the coil set. A fixed boundary single-stage method has as many degrees of freedom as stage-1 and 2 combined. It also has a very large number of competing objectives, making it challenging to find appropriate tradeoffs and challenging to find a good solution from a cold start. However, a fixed-boundary single-stage method can use the same equilibrium solvers as older two-stage methods and is therefore simpler to develop~\cite{intro_single_vmec_fixed3}.
\item \textit{Free-boundary single-stage methods}: These methods incorporate a free-boundary equilibrium solver that directly constructs the equilibrium from the magnetic field specified by the coil currents:
\begin{equation}\label{eq:free-boundary}
\begin{split}
   x'_* = \argmin_{x'}&\left[f_p(x') + f_c(x')\right], \\
        \text{subject to}\quad &g_p(x')\leq0,\quad g_c(x')\leq0.\\
        &h_p(x')=0,\quad h_c(x')=0.\\
\end{split}
\end{equation}
A free-boundary single-stage method has comparable dimensionality to a coil optimization problem and requires no matching terms. However, in general, a free-boundary equilibrium is more costly to compute than a fixed-boundary equilibrium~\cite{intro_single_overview, intro_single_spec_free, conlin2024high}. There seems to be a consensus that free-boundary stellarator solvers are challenging to converge and use in optimization, although we are unaware of strong evidence in the literature on this point. The development of accurate free-boundary solvers is an active area of study~\cite{conlin2024high}.
\end{enumerate}

\subsubsection{Quasi-single-stage (QSS) optimization}
The quasi-single-stage (QSS) method, unlike traditional single-stage methods, does not explicitly optimize the coil parameters $x'$. Instead, it attempts to improve coil-plasma balance by incorporating a coil optimization subproblem into the equilibrium optimization loop. This paper will focus on QSS using proxies based on a winding surface subproblem~\cite{qss_Yu2024-sr}. In the present notation, this is:
\begin{equation}\label{eq:QSS}
    \begin{split}
        \min_{x}&\left[f_p(x) + f_c(x'_*(x), x)\right],\\
        \text{subject to }&g_p(x)\leq0,\\
        &h_p(x)=0.
    \end{split}
\end{equation}
Here, $x_*'(x)$ is the solution to a winding surface problem. The problem must contain some form of field error objectives or constraints to ensure that $x'_*$ reproduces the equilibrium with sufficient accuracy. This serves the same purpose as $f_\text{match}(x, x')$ in \eqref{eq:fixed-boundary}. At every iteration, a winding surface solver produces a new $x_*'(x)$ and evaluates some $f_c(x'_*, x)$ as a measure of the coil complexity of equilibrium $x$.

Compared to single-stage methods, QSS methods have two key advantages. First, as discussed above, a stellarator's coils typically require more degrees of freedom to represent than the equilibrium. Therefore, a QSS method can require fewer degrees of freedom ($M\sim100$) than traditional single-stage methods ($N + M\sim 600$). Second, all existing single-stage tools are specialized for designing filament coils. In contrast, a winding surface QSS optimization can model both filamentary coils and dipole/PM arrays.

\subsection{The winding surface model}
\label{sec:theory:ws}

The winding surface model is a fast, simplified coil model. It is primarily used for initial state generation and parameter space studies. This section provides a brief introduction and review of the model.

Stellarator coil optimization, at its core, is an inverse Biot-Savart problem. The primary quantity minimized in the stellarator coil optimization problem is the normal component of the magnetic field at the plasma boundary:
\begin{equation}\label{eq:induction}
    B_\text{norm}(\x, \x') = \B(r)\cdot\hat\n = \hat\n \cdot\int_{S'} dS' \frac{I'\times(\rr-\rr')}{|\rr-\rr'|^3}.
\end{equation}
Here, $\hat\n$ is a unit normal to the plasma boundary, $S'$ is the coil surface. In this work,  $S'$ is always assumed to be a "winding surface,'' i.e., a toroidal surface with zero thickness that encloses a volume containing the entire plasma boundary. $\rr$ and $\rr'$ are locations on the plasma boundary and the coil set. $I'$ is the current distribution of the coil set. When both $I'$ and $\rr'$ are treated as unknowns, $B_\text{norm}$ is a non-convex function. However, when $\rr'$ is fixed by constraining the current to a "winding surface," $B_\text{norm}$ becomes a linear function of the unknown $I'$. This makes the minimization of the squared magnetic flux over the plasma boundary $S$, 
\begin{equation}\label{eq: least squares}
    f_B(\x, \x') = \int_S dS |B_\text{norm}|^2,
\end{equation}
a linear least-squares problem. The minimization of $f_B$ produced by an unknown, smooth surface current $\K$ is the basis of NESCOIL~\cite{merkel1987solution}, which is extensively used in the coil design of W7X and NCSX~\cite{drevlak_automated_1998, pomphrey2000compact}. Here, $\K$ is the current density distribution on the winding surface $S'$ with unit normal $\hat{n}'$. It can be defined by a scalar current potential,
\begin{align}
    \K = \hat{n}' \times \nabla \Phi'.
\end{align}
The current potential $\Phi'$ is also equivalent to the density of a dipole sheet perpendicular to the winding surface~\cite{elder_three-dimensional_2024}. 

Thanks to its mathematical simplicity, the winding surface model was popular in the early days of stellarator optimization when computational power was limited. While higher-fidelity models have become tractable, the winding surface method remains popular for initial state generation~\cite{drevlak_automated_1998},  parameter space studies~\cite{kappel_magnetic_2024, lanke_fu_global}, and designing dipole arrays~\cite{elder_three-dimensional_2024, qss_Yu2024-sr}. 

Recent improvements in winding surface methods include better conditioning, regularization, sparsity promotion, linear physics objectives, and finite element bases. Boozer~\cite{boozer_optimization_2000} performed truncated singular value decomposition (TSVD) on the induction linear operator that maps $\Phi'$ to $B_\text{norm}$ \eqref{eq:induction} to remove small-scale, high-amplitude current modes with little impact on $f_B$. In the code REGCOIL, Landreman~\cite{landreman2017improved} introduced a Tikhonov regularization term,
\begin{equation}\label{equation:cp:regcoil}
        \min_{x'}(f_B+\lambda_2f_K), \quad f_K \equiv \int_{S'} dS'|\K|^2,
\end{equation}
with a regularization weight $\lambda_2\geq0$.
This term can effectively reduce the complexity of the resulting current. Elder~\cite{elder_three-dimensional_2024} introduced L-1 regularization,
\begin{equation}\label{equation:cp:l1}
        \min_{x'}(f_B+\lambda_{1}\int_{S'} dS'\|\K\|_1),
\end{equation}
with a regularization weight $\lambda_1\geq0$. This term promotes the sparsity of the surface current. 

However, all the above formulations are unconstrained and convex optimization problems targeting the norms of $\K$. This severely limits the choice of objectives available in a winding surface model and makes it challenging to control the topology of the surface current $\K$. Our recent work on QUADCOIL aimed to address these limitations.

\subsection{QUADCOIL and coil complexity proxy}
\label{sec: theory: quadcoil}
QUADCOIL is a recent reformulation of the winding surface model~\cite{lanke_fu_global}. It enables quadratic non-convex objectives and constraints that were previously unavailable. In this section, we will briefly introduce QUADCOIL and the types of coil optimization problems that it can study.

Rather than a regularized least-squares problem, QUADCOIL formulates the winding surface problem as a non-convex quadratically constrained quadratic program (QCQP): 
\begin{equation}\label{eq:problem}
\begin{split}
    & \min_{x'} f_c(x'),\\
    &\text{subject to}\\
    &g_c(x')\leq0, \quad
    h_c(x')=0,\\
    &\text{where }f_c, g_c, h_c = \mathcal{O}[(x')^2].
\end{split}
\end{equation}
Here, $x'$ are the coil parameters, e.g., coefficients parameterizing a Fourier series expansion of $\Phi'$, along with slack variables. $f_c:\mathbb{R}^{N}\to \mathbb{R}$, $g_c:\mathbb{R}^{N}\to \mathbb{R}^{n_\text{ineq}}$, and $h_c:\mathbb{R}^{N}\to \mathbb{R}^{n_\text{eq}}$ are quadratic functions in $x'$. Here, one of $f_c$ and $g_c$ must contain some form of $B_\text{norm}$ term, such as the NESCOIL objective $f_B$. While a non-convex QCQP is NP-hard, thanks to the strong convexity of $f_B$, even when $f_c$, $g_c$, and $h_c$ include non-convex terms, the problem is still often exactly solvable as a convex semidefinite program (SDP) in polynomial-time using the Shor relaxation method~\cite{shor_quadratic_1987}. 
Notably, the existence of an exact SDP relaxation is a special property of QCQP. It does not generalize to other types of non-linear optimization problems. When $f_c$, $g_c$, and $h_c$ become generic non-linear functions, the problem can still be solved. However, the computational cost may increase, because the problem is now a general non-convex optimization problem.

By allowing non-convex quadratic penalties, QUADCOIL can study many realistic objectives that are unavailable to prior winding surface methods reviewed in Section~\ref{sec:theory:ws}. These include:
\begin{enumerate}
    \item The curvature proxy $f_\kappa^\infty$, used in~\cite{lanke_fu_global}:
\begin{equation}\label{eq:finf}
    f_\kappa^\infty\equiv\max_{S'}\|\K\cdot\nabla\K\|_\infty.
\end{equation}
    \item The self-Lorentz force~\cite{robin2022minimization}: 
    \begin{equation}
    \label{eq:ch:formulation:force}
    \begin{split}
            \frac{4\pi}{\mu_0}L'(r')= & -\oint_{S''} dS'' \frac{1}{|r'-r''|}\left\{\nabla_{r''}\cdot\left[\pi_{r''} \K(r')\right]+\pi_{r''} \K(r') \cdot \nabla_{r''}\right\} \K(r'')  \\ 
            & +\oint_{S''} dS'' [\K(r')\cdot\n(r'')] \frac{(r'-r'') \cdot\n(r'')}{|r'-r''|^3} \K(r'') \\ 
             & +\oint_{S''} dS''  \frac{1}{|r'-r''|}\left\{\K(r') \cdot \K(r'')\nabla_{r''}\cdot\pi_r''+\nabla_{r''}[\K(r') \cdot \K(r'')]\right\}\\ 
            & -\oint_{S''} dS'' [\K(r')\cdot\K(r'')] \frac{(r'-r'') \cdot\n(r'')}{|r'-r''|^3} \n(r''),\\
            \pi_r''&\equiv \mathbf I - \n(r'')\n(r'').
    \end{split}
    \end{equation}
Here, $r''$ and $r'$ are locations on the winding surface. $\pi_{r''}$ is the projection operator onto the winding surface at location $r''$. $\mathbf I$ is the identity matrix. This objective will later be used in Section\ref{sec: results: force}.
    \item  The coil-field alignment $\K\cdot \B(x')$, which is potentially useful for estimating the critical current of high-temperature superconductor (HTS) coils~\cite{formulation_IXB, formulation_IXB2}.
    \item The total stored energy in the magnetic field (implemented in filamentary optimization in Guinchard et al.~\cite{guinchard2024including}):
\begin{equation}
    E_B = \int_{\mathbb{R}^3} dV \frac{B^2}{2\mu_0}.
\end{equation}
\end{enumerate}

In addition, allowing constraints brings two benefits. First, it allows one to control the topology of $\K$ by constraining the sign of its poloidal/helical components. This is impossible with prior winding surface formulations. Second, it allows the user to directly specify a target value for each metric. In contrast, with an unconstrained winding surface code, one often needs to scan many cases with different regularization weights to obtain a solution that satisfies all engineering constraints. As we will discuss in-depth in Section~\ref{sec:numerical:benefits}, this reduces the resource usage of a QUADCOIL-based QSS optimization.

\begin{figure}
    \centering
    \includegraphics[width=0.8\linewidth]{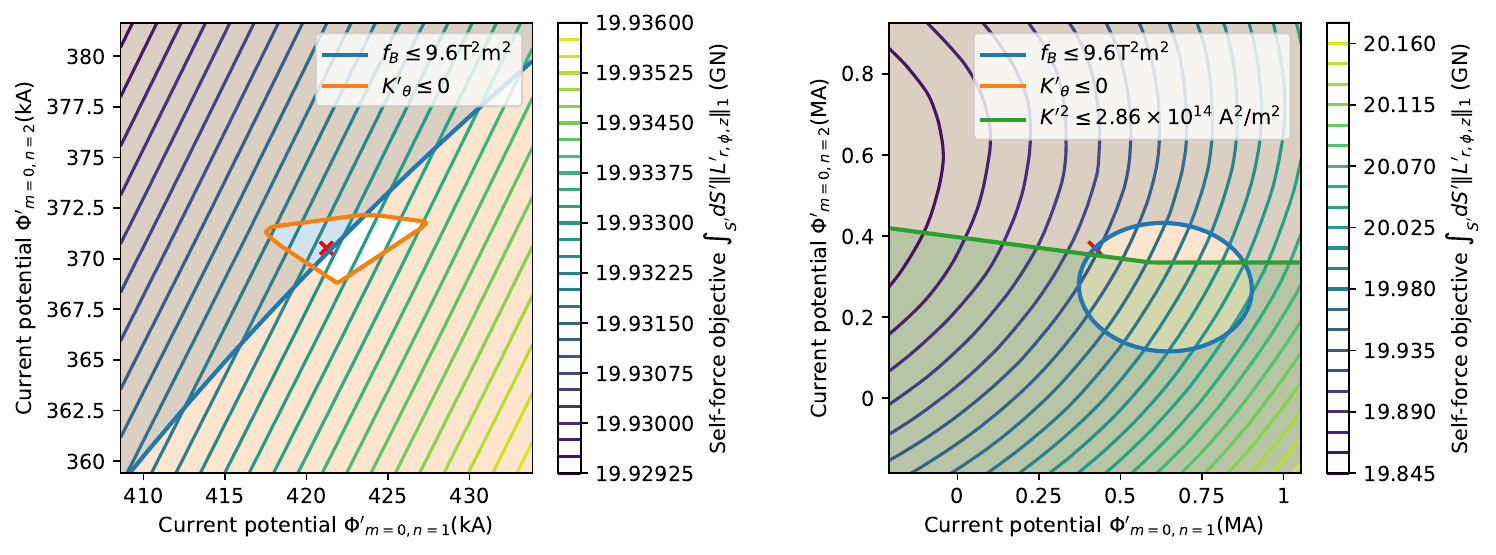}
    \caption{The objective landscape of the force minimization problem \eqref{eq: quadcoil force} near the optimum at two zoom levels. The colored areas represent regions with broken constraints. Note the apparent convexity of the problem.}
    \label{fig:landscape}
\end{figure}

As a visual illustration of the near-convexity of QUADCOIL, Fig.~\ref{fig:landscape} shows the objective landscape near the optimum of a non-convex, force-minimizing QUADCOIL problem \eqref{eq: quadcoil force}. One can see that both constraint sets exhibit convex shapes. Although the Lorentz force itself is likely non-convex, its L-1 norm, which QUADCOIL minimizes, appears convex. The choice of norms in QUADCOIL is discussed in greater detail in \cite{lanke_fu_global}. This problem will later be used to produce a low-force ARIES-CS-like equilibrium in Section~\ref{sec: results: force}. 
Therefore, we propose a QUADCOIL-based coil complexity proxy $f_c(x)\equiv f_c[x, x'_*(x)]$ for the QSS scheme. Here, $x'_*$ is the optimum of a QUADCOIL subproblem as described in Eq.~\eqref{eq:problem}. $f_c$ is the objective function of the subproblem. While one can measure metrics other than $f_c$ as coil complexity proxies, empirically, the adjoint differentiation scheme in Section \ref{sec:numerical:adjoint} works best on $f_c$.

\section{Numerical methods}
\label{sec: numerical}
This section details the numerical implementation of the differentiable QUADCOIL proxy. Section~\ref{sec:numerical:quadcoil} describes a fast, augmented Lagrangian solver for the QUADCOIL subproblem \eqref{eq:problem}. Section~\ref{sec:numerical:adjoint} describes an adjoint differentiation scheme for calculating $\nabla f_c(x)$.~\ref{sec:numerical:ws} discusses a new differentiable method for generating smooth winding surfaces. Together, these numerical methods enable the use of the QUADCOIL proxy in gradient-based optimization. As discussed in the introduction, this paper is not the first to propose QSS optimization using a winding-surface subproblem. Section~\ref{sec:numerical:benefits} discusses the advantages of QUADCOIL over the existing approaches.

\subsection{Subproblem solver}
\label{sec:numerical:quadcoil}
We choose to directly solve the non-convex QCQP \eqref{eq:problem} using the augmented Lagrangian method for speed and flexibility. Thanks to QUADCOIL's "near convexity", the speed of our augmented Lagrangian solver is comparable to the MOSEK algorithm~\cite{mosek} used in the original publication.
The augmented Lagrangian method is a popular technique for constrained optimization problems and is increasingly used for stellarator optimization problems~\cite{conlin2024stellarator,gil2025augmented}. It solves a sequence of unconstrained optimization problems,
\begin{equation}
\label{eq:ch:numerical:auglag}
    \begin{split}
        L_k(x', \lambda_k, \mu_k)= & f_c(x')+\lambda_k^\top h_c(x')+\mu_k^\top g^+_c(x', \mu_k, c_k) +\frac{1}{2} c_k\left\{\|h_c(x')\|^2_2+\left\|g^{+}_c(x', \mu_k, c_k)\right\|^2_2\right\},\\
        \text{where } (g^+_c)_j&\equiv\max\{g_j(x'), -(\mu_k)_j/c_k\}.    \end{split}
\end{equation}
Here, $\lambda_k$ and $\mu_k$ are the multipliers used at step $k$ of the augmented Lagrangian method, corresponding to the equality and inequality constraints $h_c$ and $g_c$. $c_k$ is a monotonically increasing scalar penalty factor. Each subproblem at fixed $k$ is solved using the L-BFGS method. This is a quasi-Newton algorithm commonly used for smooth, unconstrained optimization problems~\cite{lbfgs}. After each subproblem converges to a solution $x_k'$, the method updates the multipliers $\mu_k, \lambda_k$ by:
\begin{equation}
    \begin{split}
        \lambda_{k+1} & =\lambda_k+c_k h_c\left(x'_k\right), \\ \mu_{k+1} & =\mu_k+c_k g^{+}_c\left(x'_k, \mu_k, c_k\right).
    \end{split}
\end{equation}
If the solution to the inner problem, $x'_k\equiv\argmin_{x'}L_k$, satisfies the constraint to a sufficient tolerance, we update $c_k$ as follows:
\begin{equation}
    c_{k+1} = 2 c_k.
\end{equation}
The iteration continues until $\|x'_k - x'_{k-1}\|_2$ is smaller than a prescribed tolerance. It then returns the result of the final iteration as the solution $x'_*$. 

As Section~\ref{sec: numerical} shows, while it is not polynomial time, this solver is still fast enough to be integrated into an equilibrium optimization loop. A typical QUADCOIL solve requires $50-300$ total L-BFGS steps, taking $3-70$ seconds in total. This is slower than REGCOIL ($\sim0.1$ seconds)~\cite{landreman2017improved}, but still considerably faster than a typical filament optimization ($\sim20$ minutes)~\cite{lanke_fu_global}. The speed difference between QUADCOIL and REGCOIL is also less significant in an equilibrium optimization loop because a typical fixed-boundary equilibrium solve (necessary every iteration) can require tens of seconds to a few minutes~\cite{dudt2023desc}. It is also worth noting that the current implementation of the QUADCOIL proxy cold-starts the augmented Lagrangian solver in every evaluation. Future implementations will use the QUADCOIL solution from the previous equilibrium iteration as the initial guess to improve speed and accuracy.

\subsection{Adjoint differentiation of the coil subproblem}\label{sec:numerical:adjoint}
QUADCOIL uses a combination of adjoint differentiation and auto-differentiation to calculate $\nabla_xf_c(x'_*(x))$. The adjoint method (or implicit differentiation) is a method widely used to differentiate the solutions of PDEs and convex optimization problems~\cite{agrawal_differentiable_2019, lorraine_optimizing_2020, pan_bpqp_2024}. It has also gained popularity in stellarator equilibrium optimization~\cite{qss_paul_adjoint_nodate}, and was used for differentiating REGCOIL solutions in Carlton-Jones et al.~\cite{qss_carlton_regcoil}. 

The adjoint method uses the Cauchy implicit function theorem (IFT) to differentiate the solution of a PDE/optimization problem without repeated evaluations. Consider a twice-differentiable function $f:\mathbb{R}^N\times\mathbb{R}^M\to\mathbb{R}$ evaluated at a critical point $(x_0, x_0')$, $\partial f/\partial x'|_{x_0, x_0'}=0$. Then there is a small neighborhood of $(x_0, x_0')$ such that there exists a function $x'_*(x)$ such that $\partial f/\partial x'|_{x_0, x'_*(x)}=0$, and:
\begin{equation}
\label{eq:ch:QSS:ift}
    \left.\pfrac{x'_*}{x}\right|_{x_0} = -\left(\ppfrac{f}{x'}{x'}\right)^{-1}\times\left.\ppfrac{f}{x'}{x}\right|_{x_0, x'_*(x_0)}.
\end{equation}

Using the Cauchy IFT, we can now write down the derivative of the QSS objective $f_c$. We first define $S(x, z')=0$ as the stationarity condition for QUADCOIL and a new state vector of coil-related variables, $z'$, which includes $x'$ but may also contain additional information, such as slack variables and Lagrange multipliers. The forms of $S$ and $z'$ depend on the choice of solver and will be discussed in greater detail shortly. Substituting $\partial f_c/\partial x'(x, x')$ in \eqref{eq:ch:QSS:ift} with $S( x,z')$, we write
\begin{equation}
\label{eq:ift}
    \begin{split}
       \left.\pfrac{f_c(x)}{x} \right|_{x_0}
       = &\left.\pfrac{f_c(x, x')}{x}\right|_{x_0, x'(x_0)} 
       -\left.\textcolor{black}{\left.\left[ \left(\pfrac{f_c(x, x')}{z'}\right)^\top\left(\pfrac{ S(x, z')}{z'}\right)^{-1}\right]\right|_{x_0, x'_*(x_0)}}
        \textcolor{black}{\left(\frac{\partial S(x, z')}{\partial x}\right)}\right|_{x_0,x'_*(x_0)}.
    \end{split}
\end{equation}
Here, we evaluate all $x, x'$ and $z'$ derivatives using auto-differentiation. The first term $\left.\partial f_c/ \partial x\right|_{x_0, x'_*}$  is therefore trivial to evaluate. Calculating the second term requires a linear solve. The product of the second and third terms is a vector-Jacobian product (VJP). The VJP is a basic operation in auto-differentiation and does not require full knowledge of the third term. Therefore, the second term is the most costly term to evaluate, and our choice of $S$ will directly impact the accuracy and speed of adjoint differentiation. 

The choice of $S$ and $z'$ is not unique and depends on the algorithm chosen for solving \eqref{eq:problem}. This choice is a topic of active study in differentiable optimization. For example, for solvers that convert \eqref{eq:problem} into an unconstrained optimization problem, e.g., $\min_{x'} L_\text{unconstrained}(x')$, we can choose its gradient as our stationarity condition:
\begin{equation}
\label{eq:Lk_stationarity}
    z'\equiv x', \quad S(x, z') \equiv \left. \pfrac{L_\text{unconstrained}}{x'}\right|_{x, x'}.
\end{equation}
For methods where the Lagrange multipliers of $g_c, h_c$ are available, we can choose to use the KKT condition of first-order optimality for a constrained optimization problem:
\begin{gather}
    z' \equiv (x', \mu, \lambda), \quad 
    S_\text{KKT} \equiv \left[\begin{array}{c}\nabla_{x'} L(z') \\ \nabla_{\mu} L(z') \\ \nabla_{\lambda} L(z')\end{array}\right] = \left[\begin{array}{c}\nabla_{x'} L(z') \\ -g_c(x') \\ -h_c(x')\end{array}\right].\\
    L(x', \mu, \lambda)=f_c(x') - \mu^\top g_c(x') - \lambda^\top h_c(x')
\end{gather}
Here, $\mu\in\Rset^{\numineq}, \lambda\in\Rset^{\numeq}$ are the multipliers of $g_c, h_c$. $L(x', \mu, \lambda)$ is the Lagrangian. $ S_\text{KKT}$ is a rigorous definition of stationarity commonly used in differentiable convex optimization~\cite{book_nocedal_numerical_2006, agrawal_differentiable_2019, pan_bpqp_2024, qss_barratt_differentiability_2019}. However, $S_\text{KKT}$ requires accurate estimates for $\mu, \lambda$. These are not guaranteed with the augmented Lagrangian method and can be costly to obtain when strict convexity is not ensured.


Currently, QUADCOIL uses Eq.~\eqref{eq:Lk_stationarity} as the stationarity condition. Here, we choose $L_\text{unconstrained}=L_k^\text{final}$, where $L_k^\text{final}$ is the augmented Lagrangian objective at the final iteration, as defined in \eqref{eq:ch:numerical:auglag}. We treat the multipliers, $\mu$ and $\lambda$, as constants. This choice of $S$ is not theoretically rigorous, but it avoids the challenges associated with solving the full KKT system and is sufficiently accurate in practice. In Appendix~\ref{sec:appendix_penalty_barrier}, we also note how this problem can be formulated using penalty or barrier optimization techniques. However, both techniques can suffer from near singular $\partial S(x, z')/\partial z'$ near constraint thresholds~\cite{book_nocedal_numerical_2006}, which makes the second term in \eqref{eq:ift} challenging to compute.

An additional subtlety arises for continuous non-smooth objectives and point-wise constraints that are common in coil optimization. One such example is the maximum density of a dipole/PM layer, $ \max_{S'}\Phi'(x'_{*})$. For L-BFGS to converge, we must convert non-smooth problems into smooth problems. The original QUADCOIL paper achieves this by introducing slack variables \eqref{eq:ch:single-stage:thickness2}. As Appendix~\ref{sec:appendix_limitations} shows, this introduces $O(n_g'\times m_g')$ new inequality constraints and causes inaccurate gradients for $f_{c,A}$. In this paper, instead, we approximate all continuous non-smooth functions with the LogSumExp (LSE) function:
\begin{equation}
    \operatorname{LSE}(x_1, x_2, \dots, x_n)\equiv  \epsilon\log\sum_{i=1}^n\exp(x_i/\epsilon)\xrightarrow[\varepsilon \to 0]{} \max(x_1, x_2, \dots, x_n)\\
\end{equation}
Here, $\epsilon$ is an empirically chosen small constant. The LSE function is a smooth, convex approximation of the maximum. It asymptotically approaches the maximum when $\epsilon\rightarrow0$. We find that when the problem is properly normalized so that $f_c, g_c, h_c, x'\sim1$, $\epsilon=10^{-3}$ is sufficiently low to obtain good adjoint gradients for QSS optimization. Section~\ref{sec: results} will present tests of the adjoint gradient calculations versus finite-differences (Taylor tests) of realistic QUADCOIL subproblems with both non-smooth and nonconvex quantities.



\subsection{Winding surface generator}
\label{sec:numerical:ws}
Many present winding surface codes tend to generate ill-behaved surfaces with self-intersections and sharp features. Another traditional limitation of winding surface QSS is that the winding surface is fixed. Here, we describe a simple, differentiable winding surface generator designed for QUADCOIL and QSS. It produces smooth winding surfaces that can move with the plasma surface during the iteration.

\begin{figure}[htbp]
    \centering
    \begin{subfigure}[b]{0.31\textwidth}
        \includegraphics[width=\textwidth]{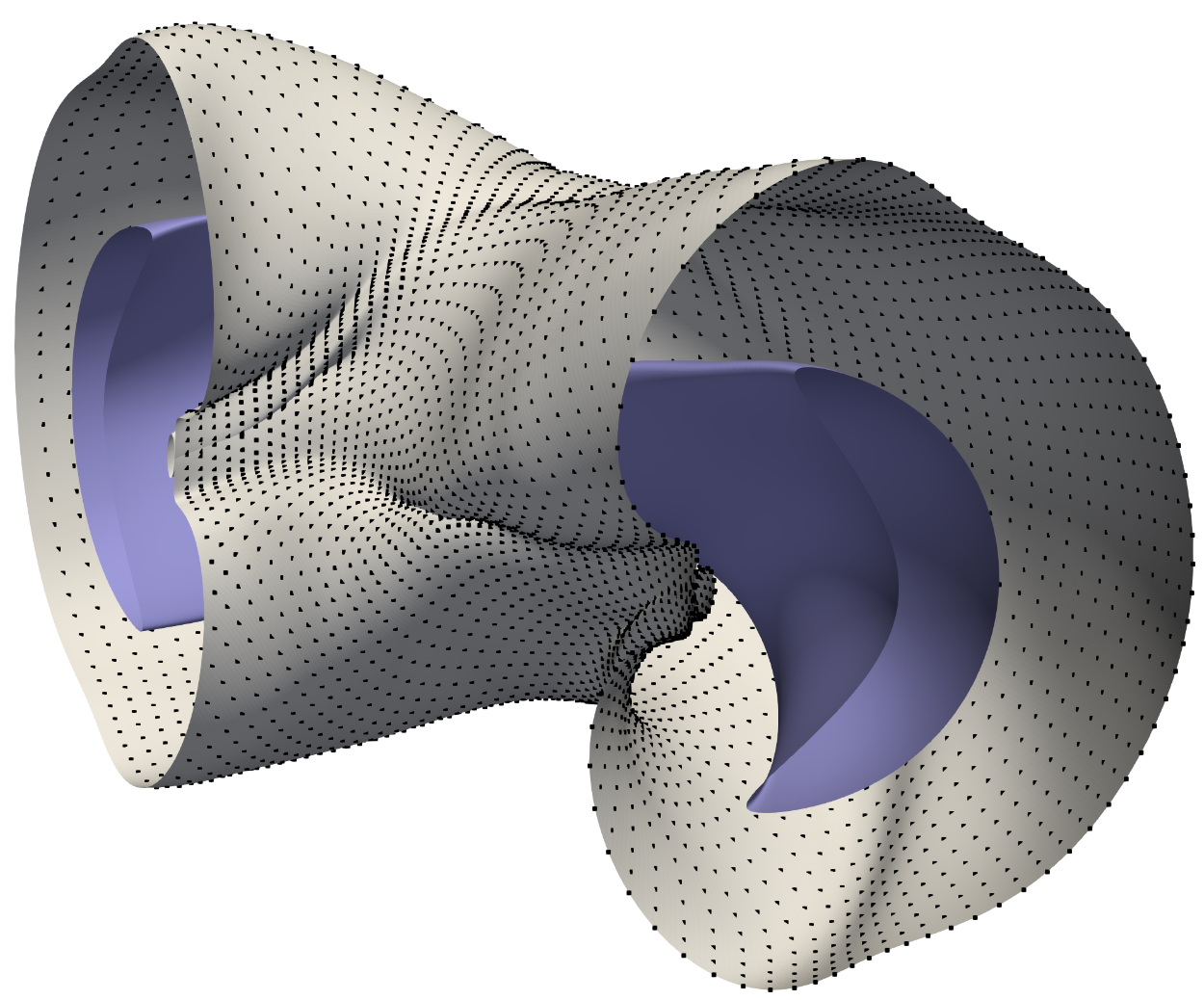}
    \end{subfigure}
    \hfill
    \begin{subfigure}[b]{0.31\textwidth}
        \includegraphics[width=\textwidth]{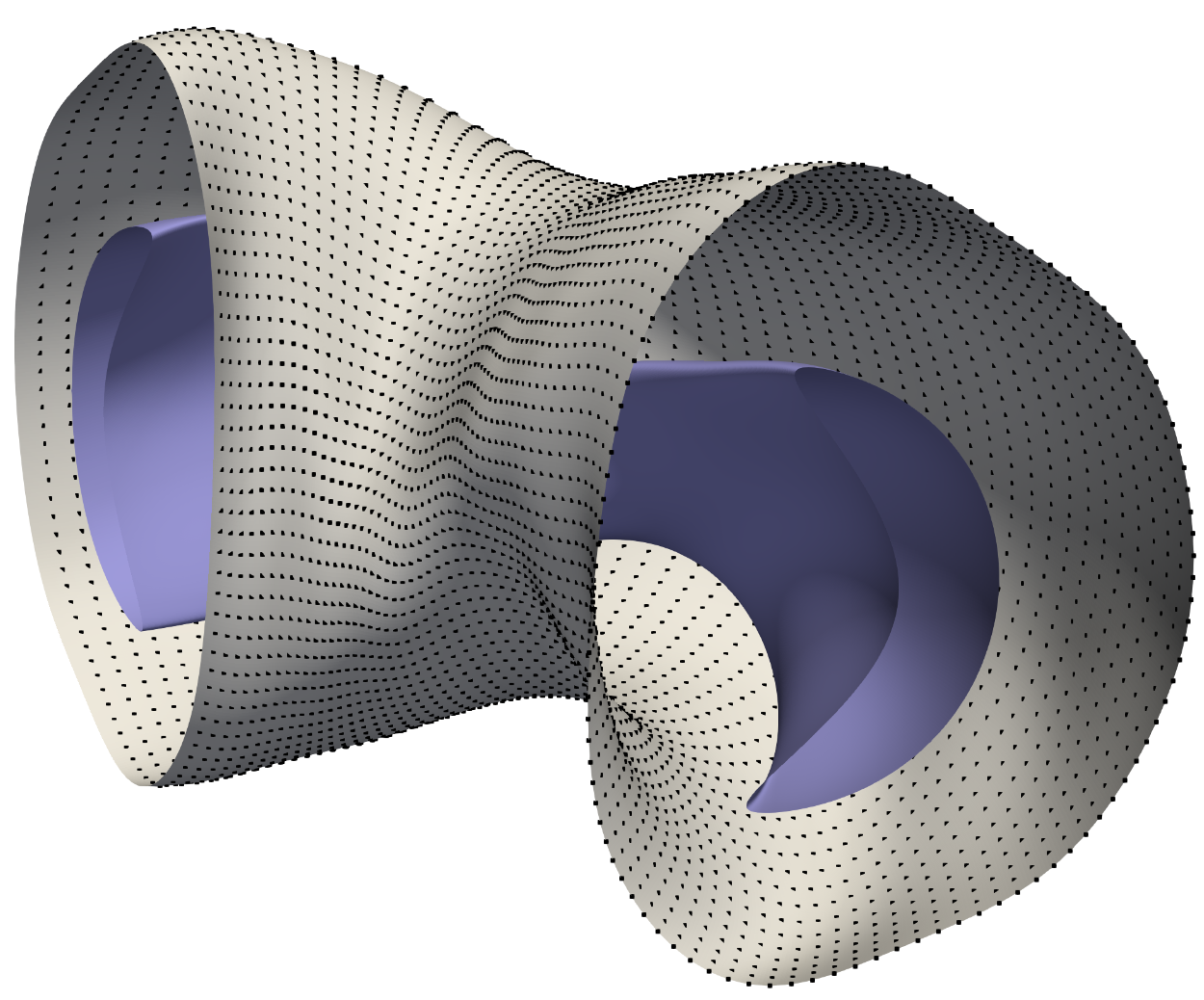}
    \end{subfigure}
    \hfill
    \begin{subfigure}[b]{0.31\textwidth}
        \includegraphics[width=\textwidth]{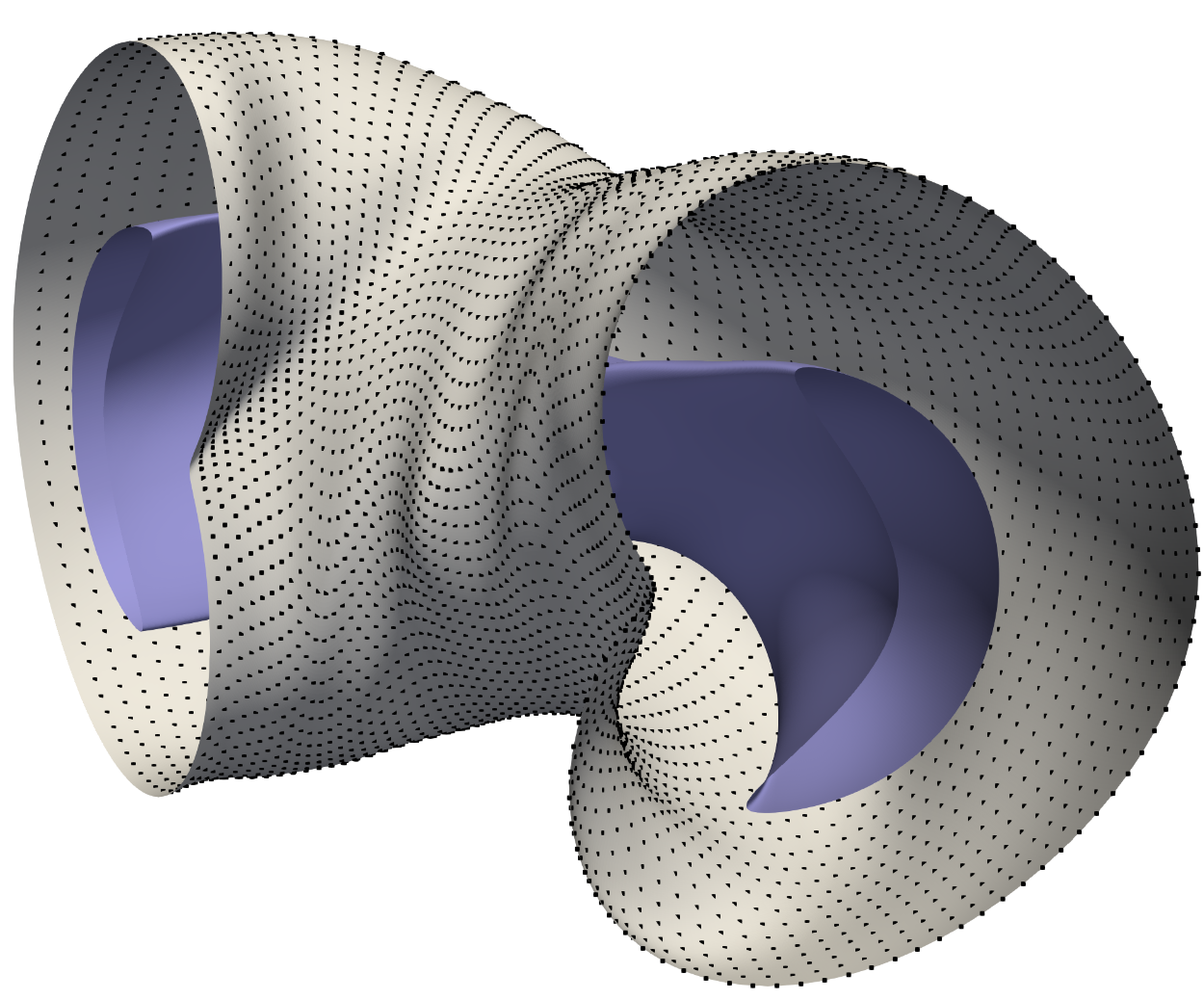}
    \end{subfigure}
    \caption{A comparison of the uniform-offset surface (left), the convex hull method (center), and Alg.~\ref{alg:ch:formulation:direct} (right) on NCSX, with $d_\text{cs}=0.5$m. Note that the new methods we present in this paper removes self-intersections, improves quadrature point uniformity, while preserving "bean-shaped" features on the inboard side.}
    \label{fig:ch:formulation:surf}
\end{figure}

Procedures for generating a well-behaved winding surface have received limited attention in the current literature. Existing winding surface codes generate the winding surface by uniformly offsetting the plasma surface along the normal direction or in the poloidal plane~\cite{landreman2017improved, kappel_magnetic_2024, landreman_simsopt:_2021}. As Fig.~\ref{fig:ch:formulation:surf} shows, when the plasma surface contains "bean-shaped" cross-sections, the uniform-offset surface often contains self-intersections. These features are especially detrimental to the minimization of the curvature proxy, $f_\kappa^\infty$ \eqref{eq:finf}, and the self-Lorentz force, $L$ \eqref{eq:ch:formulation:force}~\cite{lanke_fu_global}.

A common alternative is to iterate the winding surface shape over repeated NESCOIL/REGCOIL solves~\cite{drevlak_automated_1998, drevlak_coil_1998, pomphrey_innovations_2001}. Fundamentally, this iteration process is a nonconvex optimization problem similar to filament coil optimization. Although it has fewer degrees of freedom than filament optimization, one can argue that this approach negates the advantages of the winding surface method in speed and solution uniqueness. 

For QUADCOIL to become a valid coil complexity and force proxy, a fast, robust, and \textit{differentiable} winding surface generator is essential. The original QUADCOIL publication used a procedure that smooths each poloidal section of the winding surface by taking its convex hull. While robust, this procedure does not preserve inboard "bean-shapes" in the offset surface. It was also incompatible with auto-differentiation, as most mainstream convex hull algorithms use variable-length data formats that are not supported by JAX. Therefore, in this paper, we developed a new differentiable smoothing procedure that preserves the inboard "bean-shape" (Alg.~\ref{alg:ch:formulation:direct}).

Before discussing QUADCOIL's winding surface generators, we will first introduce some basic operations for generating winding surfaces. A common parameterization for toroidal surfaces in the stellarator literature is a Fourier expansion in the cylindrical coordinate $(r, \phi, z)$:
\begin{equation}
\label{eq:ch:formulation:surf}
\begin{split}
    r(\zeta, \theta) &= \sum_{m=0}^{m_{\text{pol}}}
    \sum_{n=-n_{\text{tor}}}^{n_\text{tor}} 
    r_{cmn} \cos(m \theta - n_{\text{fp}} n \zeta)
    + r_{smn} \sin(m \theta - n_{\text{fp}} n \zeta) ,\\
    z(\zeta, \theta) &= \sum_{m=0}^{m_{\text{pol}}}
    \sum_{n=-n_{\text{tor}}}^{n_\text{tor}}
    z_{cmn} \cos(m \theta - n_{\text{fp}} n \zeta)
    + z_{smn} \sin(m \theta - n_{\text{fp}} n \zeta).
\end{split}
\end{equation}
Here, $\zeta$ and $\theta$ are the toroidal and poloidal angles that parameterize the Fourier surface, $n$ and $m$ are the toroidal and poloidal mode numbers, $\mathcal{F}_{mn} = [r_{cmn}, r_{smn}, z_{cmn}, z_{smn}]$ are the trigonometric Fourier coefficients of the surface at fixed $(m, n)$, and $\mathcal{F}$ is the vector of all of the Fourier coefficients. We denote the operation in \eqref{eq:ch:formulation:surf} that recovers a point cloud $\rr_{kl}$ from the Fourier coefficients and uniformly sampled surface angles $\zeta_{i}, \theta_{j}$ as $\mathtt{surf\_rz\_fourier}(\mathcal{F}, \zeta_{i}, \theta_{j})$. 

A naive method for generating the winding surface for a coil set with a minimum coil-plasma distance $d_\text{cs}$ is to offset the $\numgrid = n_g \times m_g$ plasma quadrature points, $\rr_{ij} = \rr(\zeta_i, \theta_j)$, in the normal direction $\n_{ij}$:
\begin{equation}
\label{eq:ch:formulation:offset}
\begin{split}
    \rr'_{ij} & = \rr_{ij} + d_\text{cs}\n_{ij}. 
\end{split}
\end{equation}
Here, $\rr'_{ij}$ are sample points on the offset surface. Because points on the offset surface correspond exactly to points on the plasma surface, we can index points on the uniform offset surface with the same indices $i, j$ that we used for the plasma surface. 
From the point cloud $\rr'_{ij}$, we can recover the offset surface's Fourier coefficients, $\mathcal{F}'_*$, by performing a least-squares fit,
\begin{equation}
\label{eq:ch:formulation:ls}
    \mathcal{F}'_* = \argmin_{\mathcal{F}'}\sum_{i,j}[r'(\zeta'_{i}, \theta'_{j}) - r'_{ij}]^2+[z'(\zeta'_{i}, \theta'_{j}) - z'_{ij}]^2.
\end{equation}
We will denote the combined operations of Equations~\eqref{eq:ch:formulation:offset} and \eqref{eq:ch:formulation:ls} as $\mathcal{F}'_* = \mathtt{uniform\_offset}(\mathcal{F}, \zeta_i, \theta_j,d_\text{cs})$.
By construction, the uniform-offset method uses the poloidal angles $\theta_j$ that parameterize the plasma surface to parameterize the winding surface. As Fig.~\ref{fig:ch:formulation:surf} shows, this can result in uneven quadrature spacing on the winding surface. Re-parameterizing the winding surface using a new angle $\theta'_{ij}$ can improve the smoothness of the winding surface and the uniformity of the winding surface grids. Therefore, our new algorithm, Alg.~\ref{alg:ch:formulation:direct}, will use a new poloidal angle based on the arc length of each poloidal cross-section.

We are now prepared to discuss the new winding surface generator. Note that we are free to generate more quadrature points on the winding surface than on the plasma surface, so we now denote the points with new indices, $r'_{kl}$, and define $\numgrid' = n_g'\times m_g'$. Alg.~\ref{alg:ch:formulation:direct} generates a uniform offset surface and resamples $\rr'_{kl}$ at a pre-set resolution. Then, instead of taking the convex hulls, the algorithm directly finds and removes self-intersections in poloidal cross sections by following some selection rules defined through Alg.~\ref{alg:ch:formulation:rule}. 

\begin{figure}
    \centering
    \includegraphics[width=0.7\linewidth]{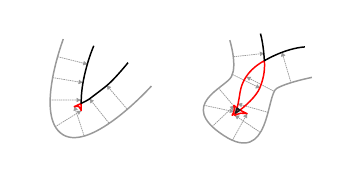}
    \caption{Limitation of Alg.~\ref{alg:ch:formulation:rule}. This figure shows two self-intersecting planar curves produced by uniform offsets. The gray curve represents the plasma surface. The figure on the right shows an example with multiple self intersections. The black curve represents the winding surface. The red portion shows parts that Alg.~\ref{alg:ch:formulation:rule} removes. Note that the routine does not work well for the complex offset curve shown on the right. Nevertheless, it works sufficiently well for the numerical studies in Section~\ref{sec: results}.}
    \label{fig:ch:formulation:caustic}
\end{figure}

Alg.~\ref{alg:ch:formulation:rule} implements a simple routine for removing self-intersections. Starting from a point on the outboard side, it performs a double loop over all line segments on each poloidal cross section. An outer loop iterates over all poloidal line segments, and an inner loop detects whether it intersects with any other segment in the same poloidal cross section. The routine generates a weight array, $w_{kl}$, for all points in $\rr'_{kl}$. Every point that is immediately followed by a poloidal line segment containing self-intersections has $w_{kl}=0$. Points that are not immediately followed by self-intersections have $w_{kl}=1$. As Fig.~\ref{fig:ch:formulation:caustic} shows, this method cannot handle highly concave plasma surfaces that cause multiple self-intersections. Fortunately, these geometries are relatively rare, and the routine works sufficiently well in our optimization. 
Finally, Alg.~\ref{alg:ch:formulation:direct}  calculates the re-parameterization $\bar{\theta}'_{kl}$. It skips sample points with $w_{kl}=0$ because dynamic-sized arrays are incompatible with JAX autodifferentiation, making it difficult to implement cubic spline smoothing. Therefore, we perform smoothing by solving a weighted, Tikhonov-regularized least-squares fit:
\begin{align}
\label{eq:ch:formulation:rls}
\mathcal{F}'_*
    \equiv&
    \argmin_{\mathcal{F}'}
    L_{\mathcal{F}'}(\rr'_{kl}, \bar\theta_{kl}, \lambda_\text{WS}, w_{kl})
    \\ \notag 
    L_{\mathcal{F}'}(\rr'_{kl}, \bar\theta_{kl}, \lambda_\text{WS}, w_{kl})
    \equiv&
    \sum_{m, n}\lambda_\text{WS} (m^2 + n^2)\|\mathcal{F}'_{mn}\|_2^2 +
    \sum_{i,j}w_{kl}[r'(\phi'_k, \bar\theta_{kl}) - r'_{kl}]^2
    +w_{kl}[z'(\phi'_k, \bar\theta_{kl}) - z'_{kl}]^2,\\\notag
    \|\mathcal{F}'_{mn}\|_2^2\equiv& r_{c,m,n}^2 + r_{s,m,n}^2 + z_{c,m,n}^2 + z_{s,m,n}^2
\end{align}
where $\lambda_\text{WS}=10^{-5}$ is the regularization parameter, and $w_{kl}$ is a weight factor individually calculated for each sample point. This way, we can simply ignore points that form self-intersections by setting their $w_{kl}=0$. The regularization term is often called the spectral density function~\cite{Henneberg2021-spectral_condensation, qss_carlton_regcoil}. This function penalizes modes with large values of $(m, n)$ that can lead to sharp features on the surface.

Fig.~\ref{fig:ch:formulation:surf} compares Alg.~\ref{alg:ch:formulation:direct} with the uniform offset method and the convex hull method used in the original QUADCOIL publication~\cite{lanke_fu_global}. The new method has two main advantages. The first advantage is that it preserves concave features on the inboard side. The second is that it does not use arrays with dynamic shapes and is therefore auto-differentiable with JAX. The main drawback is its relative lack of robustness. As mentioned above, Alg.~\ref{alg:ch:formulation:direct} can fail when the cross sections of the uniform offset surface contain complex self-intersecting geometry (Fig.~\ref{fig:ch:formulation:caustic}), whereas the convex hull-based method is robust to complex poloidal geometry due to the uniqueness of convex hulls. Alg.~\ref{alg:ch:formulation:direct}  also requires an empirical value for $\lambda_\text{WS}$. As Section~\ref{sec: results} will show, the present value works sufficiently well throughout the optimization. Nevertheless, this is not guaranteed for other equilibria. We will explore regularization methods that do not require this parameter in future research. 

\begin{algorithm}
\caption{}
\label{alg:ch:formulation:direct}
\begin{algorithmic}[1]
\State \textbf{Inputs:} plasma Fourier coefficients $\mathcal{F}$, plasma quadrature points $(\zeta_i, \theta_j)$, interpolated winding surface quadrature points $(\phi'_k, \theta'_l)$, regularization weight $\lambda_\text{WS}$, uniform offset $d_\text{cs}$.
\State $\mathcal{F}'\leftarrow\mathtt{uniform\_offset}(\mathcal{F}, \zeta_i, \theta_j, d_\text{cs})$
\Comment{Generate initial winding surface.} 
\State $\rr'_{kl}\leftarrow\mathtt{surf\_rz\_fourier}(\mathcal{F}', \phi'_k, \theta'_l)$
\Comment{Resampling.} 
\State $w_{kl}\leftarrow \mathtt{selection\_rule}(\rr'_{kl})$
\Comment{Applying selection rules.}
\State $\Delta l_{kl}\equiv \|\rr'_{kl+1} - \rr'_{kl}\|_2$
\Comment{Calculating arc-length parameterization}
\State $\bar\theta'_{kl}\leftarrow\frac{\sum_{p=1}^lw_{kp}\Delta l_{kp}}{\sum_{p=1}^{m_g'} w_{kp}\Delta l_{kp}}$
\State $\mathcal{F}'_*\leftarrow
    \argmin_{\mathcal{F}'}
    L_{\mathcal{F}'}(\rr'_{kl}, \bar\theta_{kl}, \lambda_\text{WS}, w_{kl})$.
\State \textbf{Output:} $\mathcal{F}'_*$
\Comment{Winding surface coefficients.}
\end{algorithmic}
\end{algorithm}

\begin{algorithm}
\caption{\texttt{selection\_rule}}
\label{alg:ch:formulation:rule}
\begin{algorithmic}[1]
\State \textbf{Input:} $\rr'_{kl}$
\Comment{Quadrature points.}
\For{$k = 1$ to $n_g$}
    \Comment{Looping over poloidal contours}
    \State $c\leftarrow1$ 
    \Comment{A carry variable.}
    \State Assume $\rr'_{1l}$ is not a part of any self-intersections. 
    \For{$l = 1$ to $m_g$}
    \Comment{Looping over vertices}
        \For{$p = 1$ to $m_g$}
            \Comment{Detecting if the cross section intersects segment $l$}
            \If{$(\rr'_{kl}, \rr'_{k(l+1)}), (\rr'_{kp}, \rr'_{k(p+1)})\,\mathbf{intersect}$}
                \State $c\leftarrow 1-c$ 
                \Comment{Flipping $c$ between 0/1}
                \State \textbf{break}
            \EndIf
        \EndFor
        \State $w_{kl}\leftarrow c$
        \Comment{$w_{kl}$ flips 0/1 at self-intersections.}
    \EndFor
\EndFor
\State \textbf{Output:} $w_{kl}$
\Comment{Weights based on self-intersection.}
\end{algorithmic}
\end{algorithm}

\subsection{Comparison with existing works}
\label{sec:numerical:benefits}
This paper is not the first to propose QSS optimization using a winding surface subproblem. In 2023, Yu et al. presented MUSE++, a MUSE-like equilibrium optimized with a REGCOIL-based coil complexity proxy~\cite{qss_Yu2024-sr}. Because REGCOIL is unconstrained, Yu's approach performs 30 REGCOIL solves that scan the regularization weight $\lambda_2$ (see \eqref{equation:cp:regcoil}):
\begin{equation}
    x'_i=\argmin_{x'}\left[f_B + \lambda_{2,i} f_K\right], \quad
    \lambda _{2, i} = 10^{-13}, ..., 10^{-24}\,\,(\text{T}^2\text{m}^2/\text{A}^2)
\end{equation}
Here, $\lambda_{2, i}$ are the values of $\lambda_2$ covered in the REGCOIL parameter scan. The scan points are fixed throughout the QSS optimization. After performing all 30 REGCOIL solves for $x_i'$, it chooses one solution, $x'_{i_*}$, by finding the index $i_*$ that minimizes the product of $f_B$ and $f_K$:
\begin{equation}
      i_* = \argmin_{i} f_B(x'_i)f_K(x'_i).
\end{equation}
This index marks the scan point closest to the "inflection point" on a plot of $f_B$ and $f_K$. The QSS objective is a weighted sum of the $\Phi'_\text{max}$ and $f_B$ measured at the inflection point:
\begin{align}
    f_\text{c, Yu}(x) &= w_\Phi[\Phi'^2_\text{max}(x, x'_{i_*})] + w_Bf_B(x, x'_{i_*}).
\end{align}
Here, the weights $w_\Phi$ and $w_B$ are chosen empirically.

We build on Yu's innovative quasi-single-stage method by replacing the REGCOIL parameter scan with a more realistic QUADCOIL winding surface subproblem. REGCOIL can only target the weighted sum of $f_K$ and $f_B$, while QUADCOIL supports more realistic objectives and constraints. The QUADCOIL subproblem can also be configured to reflect specific coil engineering requirements for different stellarator devices. As Section~\ref{sec: results: PM} shows, the QUADCOIL QSS optimization converges in fewer iterations and produces equilibrium with simpler PM arrays than MUSE++. Section~\ref{sec: results: force} shows that the QUADCOIL proxy can simplify filament coils as well.

There are other analytic coil complexity proxies for stage-1 optimization that are not based on winding surface subproblems. The most notable example is the L-grad-B proxy, which has been shown to correlate with coil-plasma distances~\cite{kappel_magnetic_2024}. Compared to a winding surface subproblem, analytic proxies are simpler to implement and less expensive to evaluate. However, different coil engineering requirements may favor different plasma properties. A key advantage of QUADCOIL over analytic proxies is its ability to model coil sets with specific engineering requirements.

\section{Numerical results}
\label{sec: results}
This section presents two numerical studies to validate the effectiveness of the QUADCOIL proxy. The first study is designed to minimize the current dipole density, and the second study is designed to minimize the coil forces. In both studies, we (1) perform QSS optimization, starting from a known equilibrium; (2) discard the winding surface solution; (3) design permanent magnets or filamentary coils from scratch for both equilibria; (4) compare their performance. In both studies, coils of the QSS equilibrium outperforms that of the original equilibria. Additional benefits could be obtained by cutting coils or otherwise initializing from the optimized winding surface, but we found in practice that keeping only the plasma surface works quite well. 


Both studies solve the following QSS problem:
\begin{equation}\label{eq:quasi-single example}
\begin{gathered}
        \min_x \left\{
            30[V(x) - V_\text{init}]^2 
            + 30[\iota_{\text{axis}}(x) - \iota_{\text{axis, init}}]^2 
            +30[\iota_{\text{edge}}(x) - \iota_{\text{edge, init}}]^2 
            +500\hat f_\text{T}(x) 
            +10f_c(x)\right\},\\
            r_{c00} = (r_{c00})_\text{init} ,\psi_\text{LCFS} = \left(\psi_\text{LCFS}\right)_\text{init}, \\
            \text{(if finite $\beta$) } p(\psi) = p_\text{init}(\psi), \iota(\psi)=\iota_\text{init}(\psi)
\end{gathered}
\end{equation}
Here, $r_{c00}$ is the coefficient of the zeroth-radial-harmonic of the plasma boundary. This is a measure of the configuration's major radius. $\psi_\text{LCFS}$ is the total toroidal flux, and $V(x)$ is the total plasma volume. The terms subscripted with "init'' are prescribed target values, the choice of which depends on the desired plasma surface. The first three terms in \eqref{eq:quasi-single example} are common terms in conventional stage-1 optimization~\cite{imbert-gerard_introduction_2024}. Along with the constraints, these terms ensure that the plasma parameters of the resulting equilibrium match those of the initial state. $\hat f_\text{T}$ is a triple-product QS metric, given by~\cite{dudt2023desc}:
\begin{gather}
    \hat f_\text{T}=\int_\text{V} d^3V \max\left[\frac{\langle R\rangle^2|f_\text{T}|}{\langle B\rangle^4}
    - \left(\frac{\langle R\rangle^2|f_\text{T}|}{\langle B\rangle^4}\right)_\text{init}, 0\right]^2,\\
    f_\text{T}\equiv\nabla \psi \times \nabla B \cdot \nabla (B \cdot \nabla B),
\end{gather}
where $\langle R \rangle$ is the effective major radius, and $\langle B \rangle$ is the average magnetic field strength. The last term $f_c(x)$ is the QUADCOIL proxy, which we will choose based on particular features we want to promote during coil design. The goal of \eqref{eq:quasi-single example} is to achieve point-wise QS that is at least as good as the initial configuration and to use any additional slack in the problem to minimize the QUADCOIL proxy (obtain feasible coils). As we will see in the following sections, by choosing $f_c, g_c$ and $h_c$, the QUADCOIL proxy can flexibly model a variety of coil design problems. 

The QSS optimization is conducted in DESC~\cite{dudt2023desc, dudt_desc_2020} using the trust-region algorithm. Both studies use 8 toroidal and 17 poloidal Fourier modes for the current potential $\Phi'$. When optimizing the plasma, it is well known that "Fourier continuation'', solving a sequence of subproblems with an increasing number of Fourier modes, greatly improves performance~\cite{landreman2021stellarator,conlin2024stellarator,jang2025exponential}. The QSS method does not alter this fact. Therefore, we perform the first QSS optimization in $5$ Fourier continuation steps, with $48 - 194$ degrees of freedom; the second QSS in $6$ Fourier continuation steps, with $48-186$ degrees of freedom. As discussed in Section~\ref{sec: theory}, both QSS problems have substantially fewer parameters than a typical single-stage optimization.

\subsection{Sparse PM array}
\label{sec: results: PM}
MUSE is an operational two-field-period, stellarator symmetric table-top experiment that shapes its 3D magnetic fields using a set of simple, axisymmetric toroidal field coils and a large array of thousands of permanent magnets~\cite{qian2022simpler,qian2023design}. Using the QSS method, Yu et al.~\cite{qss_Yu2024-sr} designed a new MUSE++ plasma equilibrium with substantially improved quasi-axisymmetry. We will imitate their problem setup for a direct comparison between methods. The winding surface used in this problem is a fixed axisymmetric toroidal shell with a major radius of $0.3$m and a minor radius of $0.1$m. This is roughly modeled after the dimensions of the permanent magnet holder in the MUSE device. After obtaining a QSS solution, the winding surface is removed (only the new plasma surface is retained) and a permanent magnet optimization is then performed on the true permanent magnet grid used by the MUSE experiment. We show in this section that QUADCOIL QSS optimization can find a new MUSE-like plasma boundary that exhibits very similar plasma performance but requires only $\approx 65\%$ of the permanent magnets needed for the original MUSE plasma surface. As a further improvement over existing works, we use a realistic model for the "L-2'' planar coils used in the MUSE experiment~\cite{intro_muse}, rather than an infinite straight wire~\cite{qss_Yu2024-sr} or a uniform poloidal net current~\cite{lanke_fu_global, landreman2017improved, merkel1987solution}. The shape and currents of the planar coils are fixed during the QSS optimization.

To demonstrate our capabilities, we generate two MUSE-like vacuum fields using QUADCOIL proxies that minimize PM layer thickness and count, subject to a constraint of sufficient field accuracy. The first new vacuum field (A) uses a proxy that minimizes the peak density of a dipole sheet:
\begin{equation}\label{eq: quadcoil A}
    \begin{split}
    f_{c,A}(x)= \max_{S'}\Phi'(x'_{*}),\quad
        x'_* = &\argmin_{x'} \left(\max_{S'}\Phi'\right)\\
        &\text{subject to}\\
        &f_B\leq 10^{-6}\,\text{T}^2 \text{m}^2.
    \end{split}
\end{equation}

\begin{figure}
    \centering
    \includegraphics[width=0.75\linewidth]{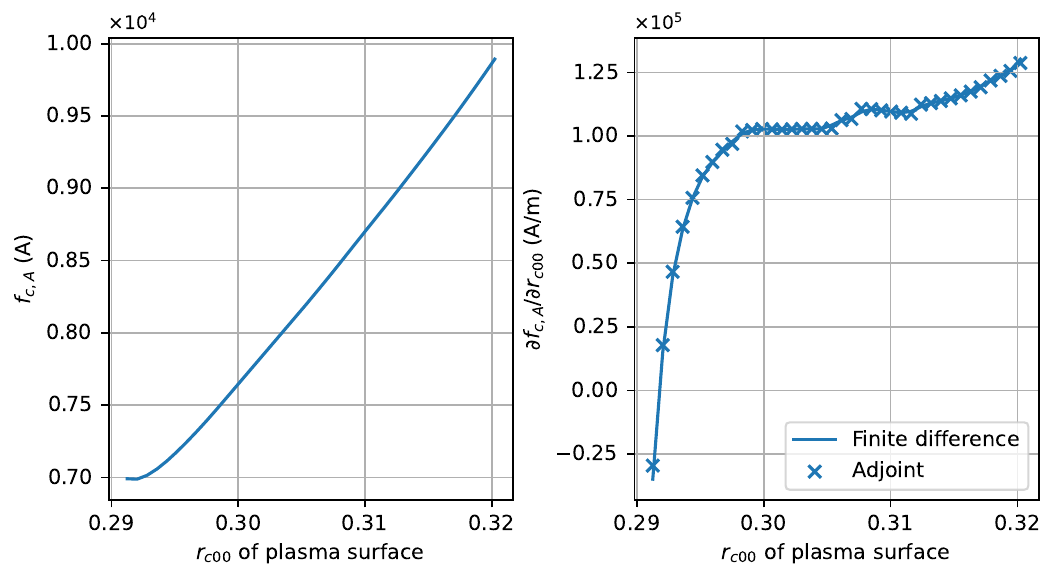}\\
    \includegraphics[width=0.75\linewidth]{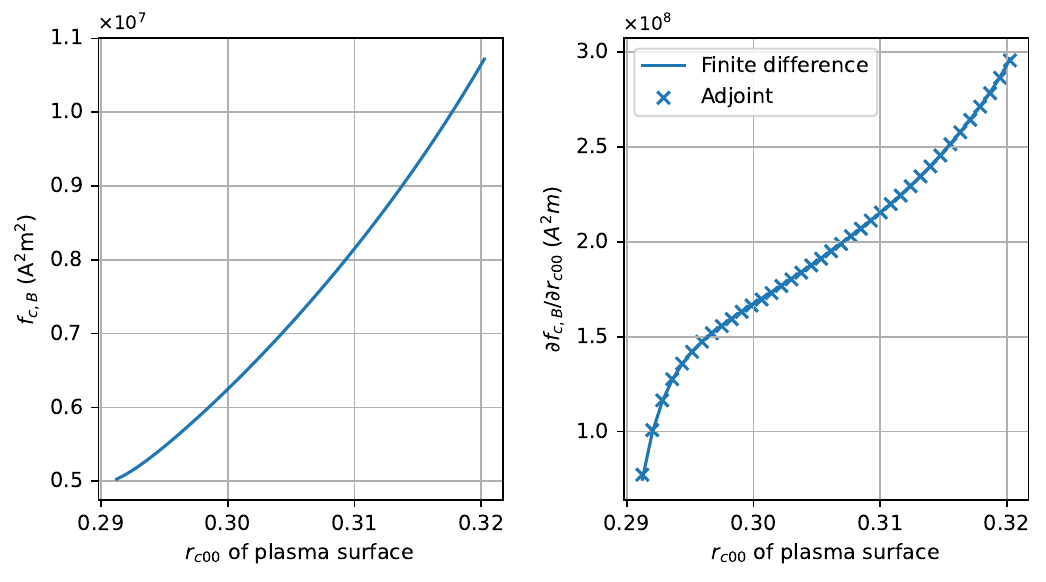}\\
    \caption{The values (left) and partial derivatives (right) of $f_{c, A}$ and $f_{c, B}$ w.r.t. the plasma Fourier coefficient $r_{c00}$. The finite difference grid size is $\approx10^{-3}r_{c00}$. Note the close agreement between the finite difference and the adjoint gradients.}
    \label{fig:ch:single-stage:taylor}
\end{figure}

The second new vacuum field, (B), uses a proxy that minimizes the sum of squared dipole density:
\begin{equation}\label{eq: quadcoil B}
    \begin{split}
        f_{c, B}(x) =  \frac{1}{2}\int_{S'}dS'\,|\Phi'(x'_*)|^2,\quad
        x'_* = &\argmin_{x'} \frac{1}{2}\int_{S'}dS'\,|\Phi'|^2\\
        &\text{subject to}\\
        &f_B\leq 10^{-6}\,\text{T}^2 \text{m}^2.
    \end{split}
\end{equation}
Fig.~\ref{fig:ch:single-stage:taylor} shows the Taylor tests for $f_{c,A}(x)$ and $f_{c,B}(x)$. Note the close agreement between the adjoint and finite difference gradients. 

\begin{figure}
    \centering
    \includegraphics[width=1\linewidth]{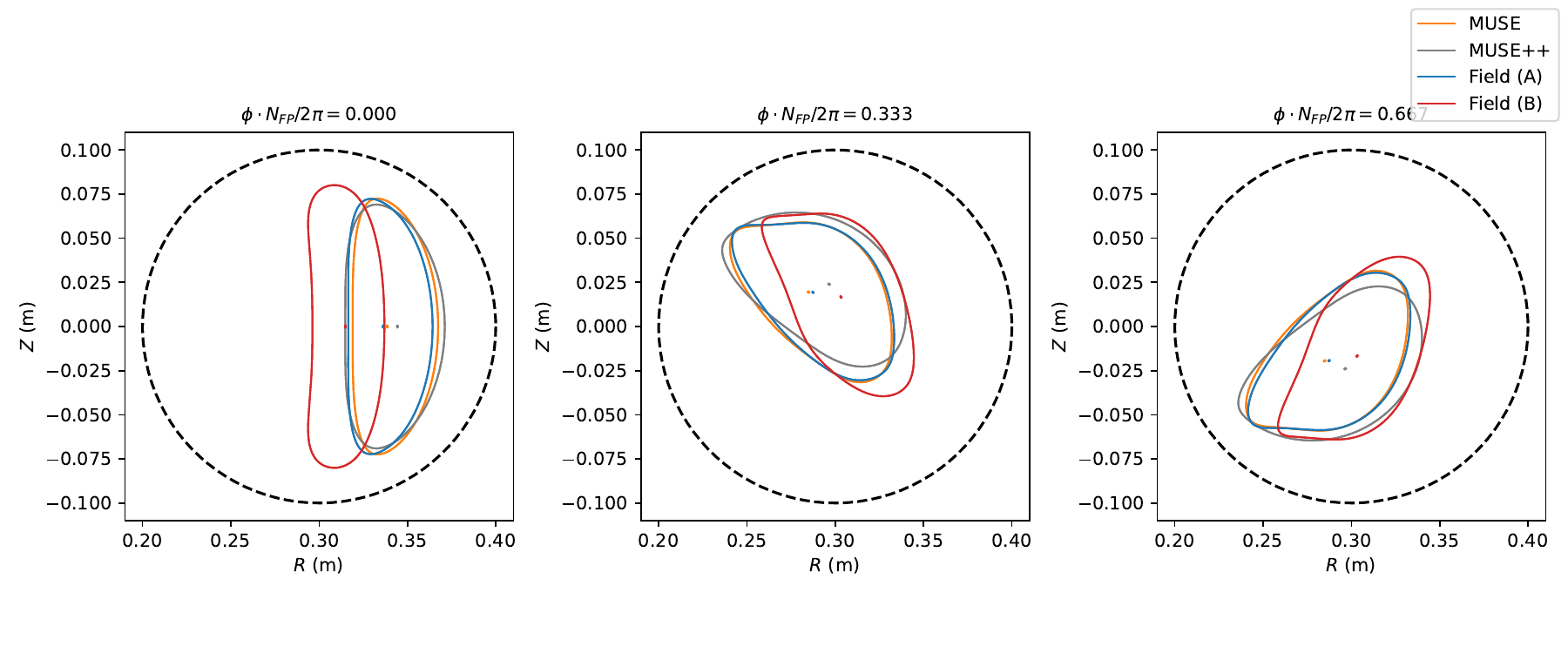}
    \caption{Plasma boundaries and axes of MUSE, MUSE++, (A) and (B). The dashed lines represent the PM holder, which we use as the winding surface. Note that compared to (A)'s, (B)'s boundary deviates further from that of MUSE.}
    \label{fig:ch:single-stage:flux}
\end{figure}

\begin{figure}
    \centering
    \includegraphics[width=0.85\linewidth]{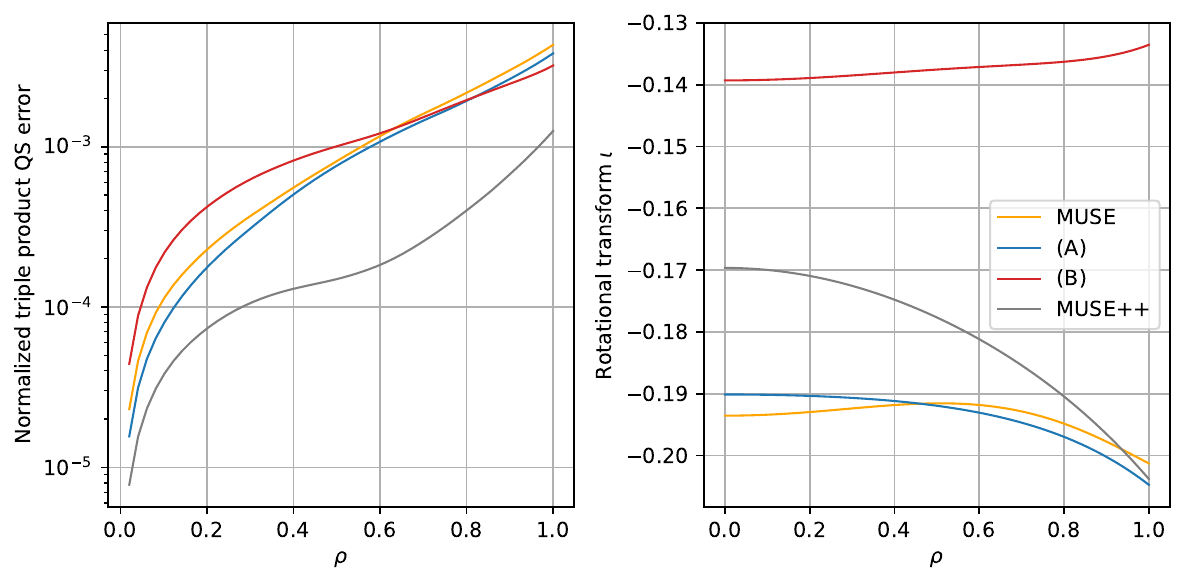}

    \caption{The QS quality values (left) and rotational transform (right) of MUSE, MUSE++ and the new vacuum fields. Here, $\rho\equiv\sqrt{\psi/\psi_\text{LCFS}}$ is the radial coordinate used by DESC.}
    \label{fig:ch:single-stage:qs}
\end{figure}

Fig.~\ref{fig:ch:single-stage:flux} compares the outer flux surfaces of MUSE, MUSE++, and the new vacuum fields. Fig.~\ref{fig:ch:single-stage:qs} compares the rotational transform and QS quality of MUSE, MUSE++, and the new vacuum fields. Both (A) and (B) have QS quality comparable to MUSE but worse than MUSE++. This is expected, as the QS quality term $\hat f_\text{T}$ in \eqref{eq:quasi-single example} will only attempt to match the QS quality of MUSE. As Appendix~\ref{sec:appendix_musepp} shows, while it is possible to achieve QS quality comparable to MUSE++ using QUADCOIL QSS optimization, the optimization terminates without making significant changes to the plasma boundary, yielding marginal improvements in $f_{c,A}$ and $f_{c, B}$. This suggests that MUSE++ may be a narrow local minimum in the QS objective landscape. 

\begin{figure}
    \centering
    \includegraphics[width=0.8\linewidth]{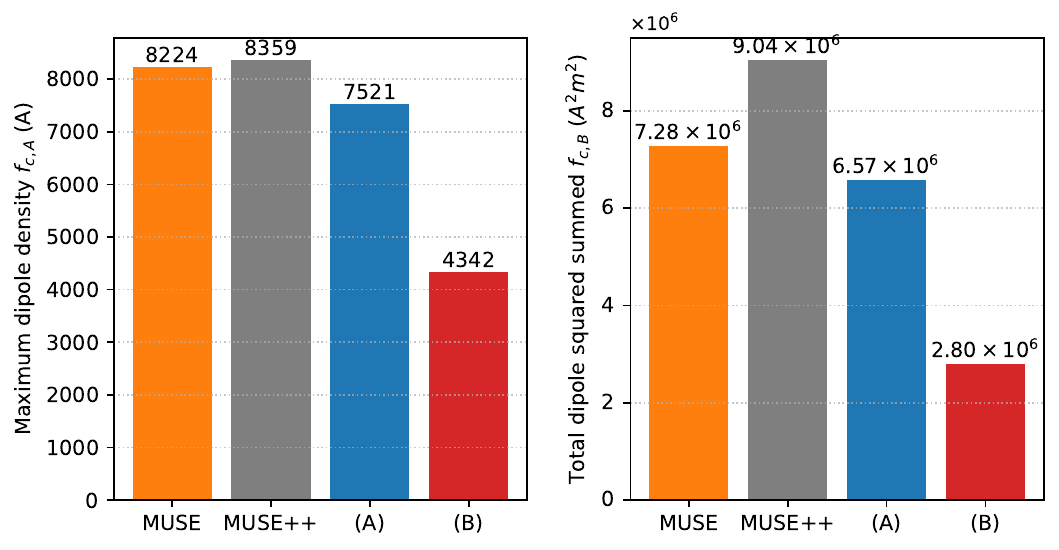}
    \vspace{1em}
    \caption{Comparison of the dipole thickness and count among MUSE, MUSE++, and the two new vacuum fields.}
    \label{fig:bar}
\end{figure}

Figure~\ref{fig:bar} and Table \ref{tab:proxy} compare the values of $f_{c, A}$ and $f_{c, B}$ among the four vacuum fields. As the figure shows, both (A) and (B) outperform MUSE and MUSE++ in $f_{c, A}$ and $f_{c, B}$. This is achieved without significant degradation in QS quality or rotational transform compared to MUSE. Vacuum field (B) outperforms (A) in both $f_{c, A}$ and $f_{c, B}$, suggesting that the landscape of $f_{c, B}$ may be smoother. It also seems likely that case (B) performs very well by "cheating'' on the target for the rotational transform (it does not match the MUSE rotational transform well in Fig.~\ref{fig:ch:single-stage:qs}), and in doing so, finds strong improvements in the other metrics.

\begin{table}[]
    \centering
       \begin{tabular}{ccccc}\toprule
         &  MUSE&  MUSE++&  Vacuum field (A)& Vacuum field (B)\\\midrule
         \makecell{Maximum dipole\\density $f_{c,A}$ (A)}&  $8224$&  $8359(+1.6\%)$&  $7521(-8.5\%)$& $4342(-47.2\%)$\\
         \makecell{Dipole count\\$f_{c,B}$(A$\text{m}^2$)}&  $7.28\times10^6$&  $9.04\times10^6(+24.1\%)$&  $6.57\times10^6(-9.7\%)$& $2.80\times10^6(-61.5\%)$\\ \bottomrule
    \end{tabular}
    \caption{Comparison of the QUADCOIL proxy of MUSE++, (A) and (B) with the MUSE values.}
    \label{tab:proxy}
\end{table}

Tab.~\ref{tab:time} shows a comparison between the computational resources and time required to produce MUSE++, (A) and (B). While each QUADCOIL evaluation is slower than REGCOIL, the QUADCOIL QSS optimization converges in substantially fewer iterations and in a shorter time. The QUADCOIL QSS optimization also requires fewer computational resources, running on a single RTX 4060 (8GB) GPU instead of 16-CPUs. This is likely due to a combination of factors: QUADCOIL's adjoint differentiation scheme may be more robust than REGCOIL finite differencing. DESC uses auto-differentiation and may be able to compute gradients with lower resource use and better accuracy than VMEC finite differencing. QUADCOIL and DESC are both based on JAX \cite{jax2018github}. JAX performs just-in-time compilation and optimizes code structure for execution on GPUs. This may have allowed both codes to be compiled together and run in parallel on one GPU. 

\begin{table}
    \centering
    \begin{tabular}{cccc}\toprule
         &  MUSE++&  Vacuum field (A)& Vacuum field (B)\\\midrule
         Total time&  $12$hrs, $16$ CPUs&  $40$mins, RTX 4060& $11.5$ minutes, RTX 4060\\
 Total iterations& $\approx 3000$& $34$&$57$\\
 Time per iteration& $\approx12$s& $70.7$ s&$12.1$ s\\
 Time per function evaluation& Not published& $15$ s&$3.69$ s\\ \bottomrule
    \end{tabular}
    \caption{Resource and run time statistics of MUSE++, field (A) and field (B).}
    \label{tab:time}
\end{table}

\begin{figure}
    \centering
    \includegraphics[width=0.75\linewidth]{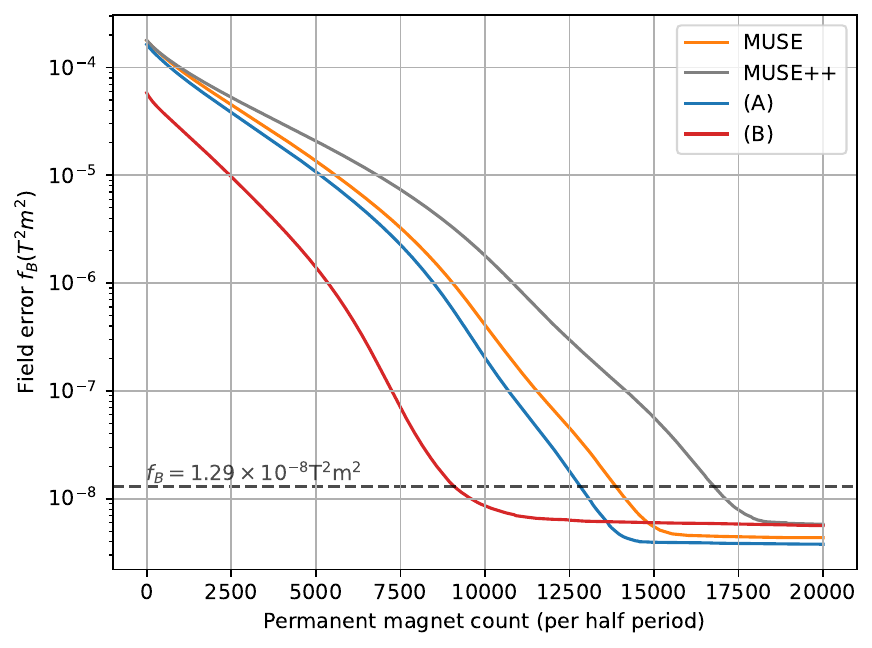}
    \caption{The accuracy vs. number of magnets tradeoff of GPMO on MUSE, MUSE++, (A) and (B). The lower left corner represents lower field error and dipole counts. The dashed line represents the $f_B$ value in the MUSE final design.}
    \label{fig:pareto_gpmo}
\end{figure}

After obtaining promising QSS results, we remove the winding surface and retain the optimized plasma surfaces. We then perform discrete PM optimization on MUSE, MUSE++, (A) and (B) using the greedy permanent optimization algorithm (GPMO)~\cite{kaptanoglu_greedy_2023,hammond2024improved,ulrich2025permanent} on the true MUSE permanent magnet arrays. Note that the original MUSE and MUSE++ PM configurations are produced with a different algorithm (FAMUS) \cite{intro_pm_opt}. Although FAMUS outperforms GPMO at large dipole-counts and small $f_B\sim10^{-9}\text{T}^2\text{m}^2$, we choose GPMO in this study for its finer control in PM counts and final $f_B$ values. This enables the fine scan of over $f_B$ in the following paragraph, which would be challenging with FAMUS. Although it is possible to initialize the discrete PM optimization with the QUADCOIL solutions to potentially accelerate convergence, we cold start all PM optimizations for fairness. Fig.~\ref{fig:pareto_gpmo} shows the tradeoff between the number of magnets and the field accuracy of the PM optimizations in all four vacuum fields. After GPMO, MUSE++ consistently has higher field errors and PM counts than MUSE. On the other hand, both (A) and (B) consistently achieve lower field errors and PM counts than MUSE and MUSE++. 

Tab.~\ref{tab:gpmo} shows the four GPMO solutions with $f_B$ values matching that of the MUSE final design ($1.29\times10^{-8}\text{T}^2\text{m}^2$). Among all cases, field (B) consistently outperforms the rest in a large range of $f_B$ . Compared to MUSE's GPMO solution, (B)'s solution achieves comparable magnetic field accuracy with $34.2\%$ fewer PMs. This confirms that the QUADCOIL proxy is indeed effective in reducing an equilibrium's PM/dipole array complexity. In some ways, this is a surprising result; we have completely discarded the winding surface solutions from the QSS optimization, and the "memory'' of the reduced dipole count on the winding surface is only encoded in the new plasma surface. This memory is apparently sufficient to find significant performance gains in a subsequent full-scale permanent magnet optimization. 

\begin{figure}[htbp]
    \centering
    \begin{subfigure}[b]{0.36\textwidth}
        \includegraphics[width=\textwidth]{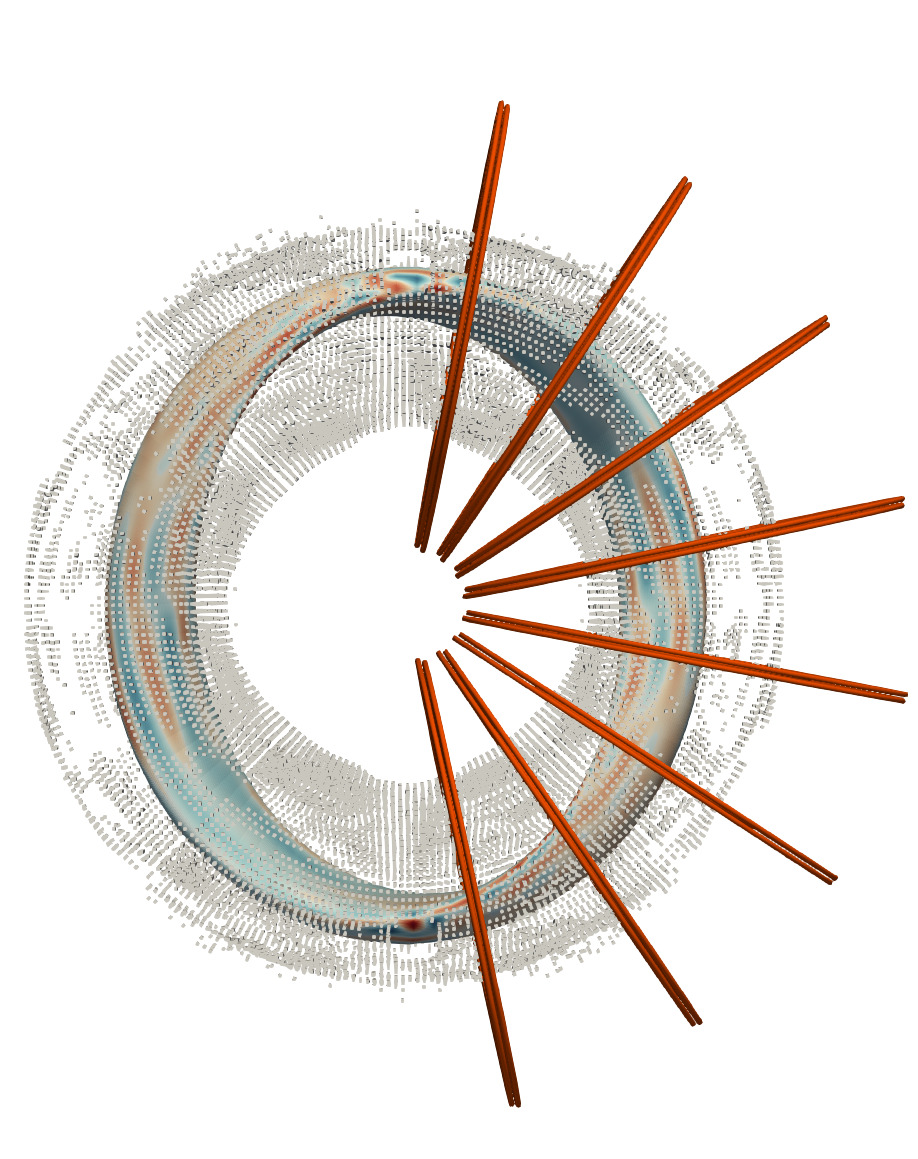}
        \caption{Vacuum field (A)}
    \end{subfigure}
    \begin{subfigure}[b]{0.44\textwidth}
        \includegraphics[width=\textwidth]{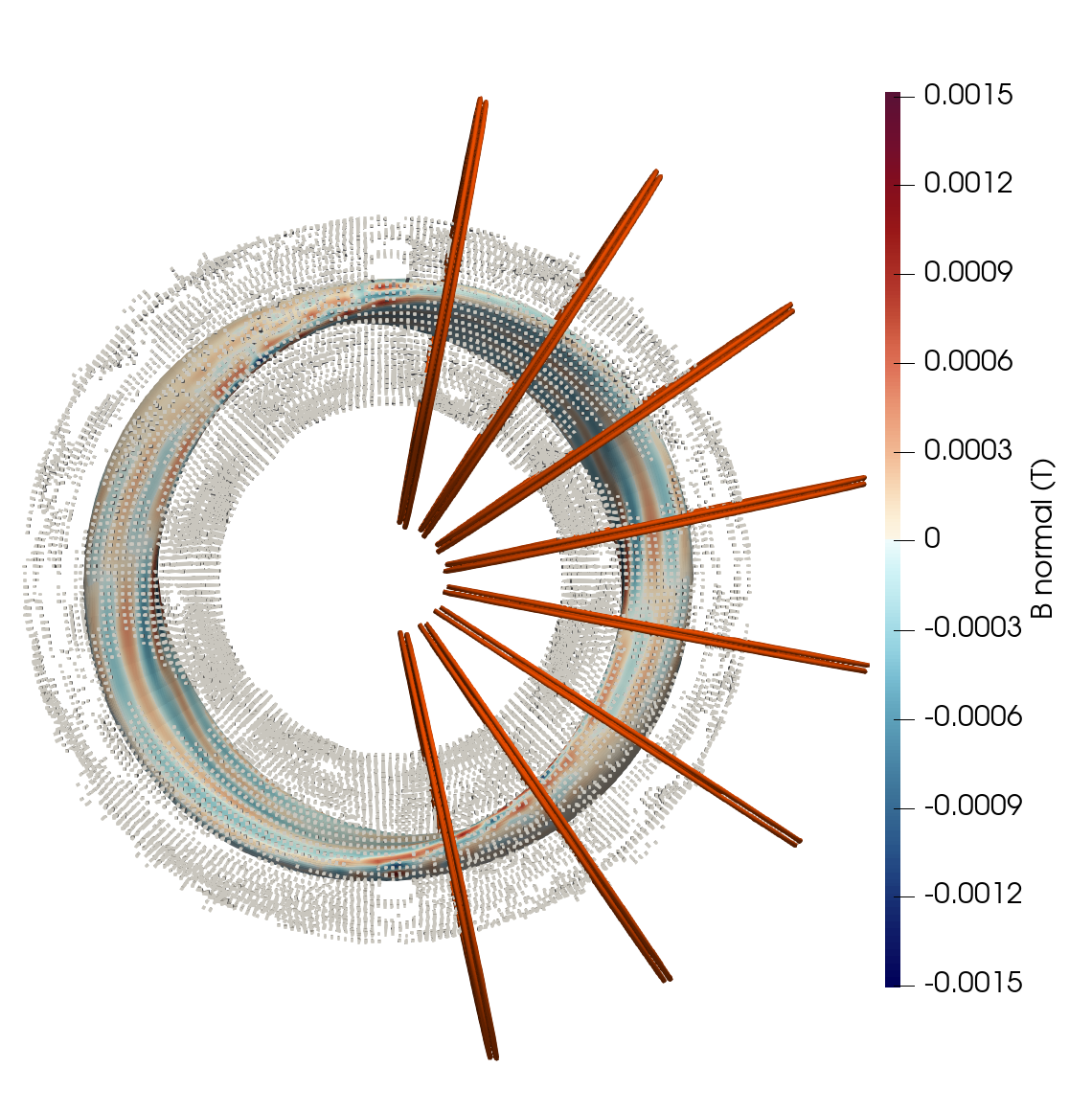}
        \caption{Vacuum field (B)}
    \end{subfigure}\\
    \centering
    \begin{subfigure}[b]{0.36\textwidth}
        \includegraphics[width=\textwidth]{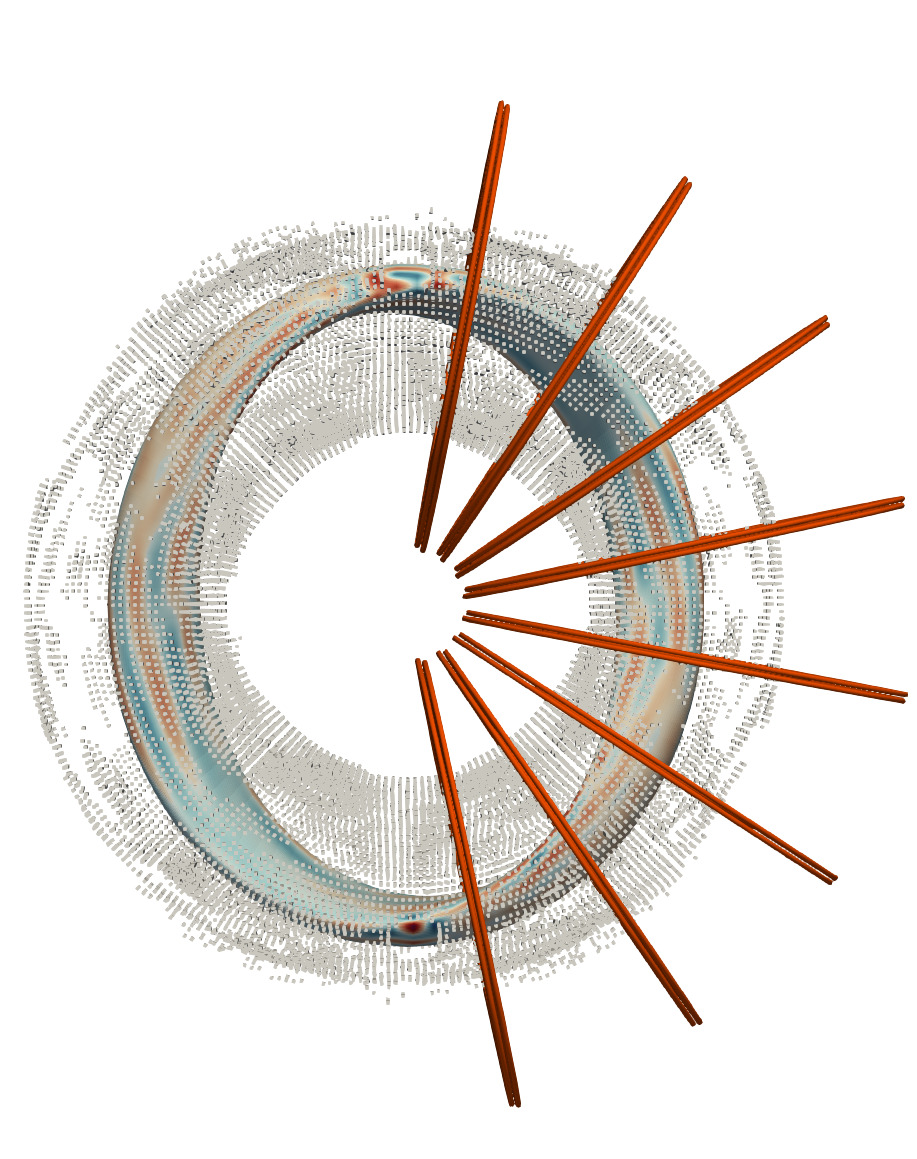}
        \caption{MUSE}
    \end{subfigure}
    \begin{subfigure}[b]{0.44\textwidth}
        \includegraphics[width=\textwidth]{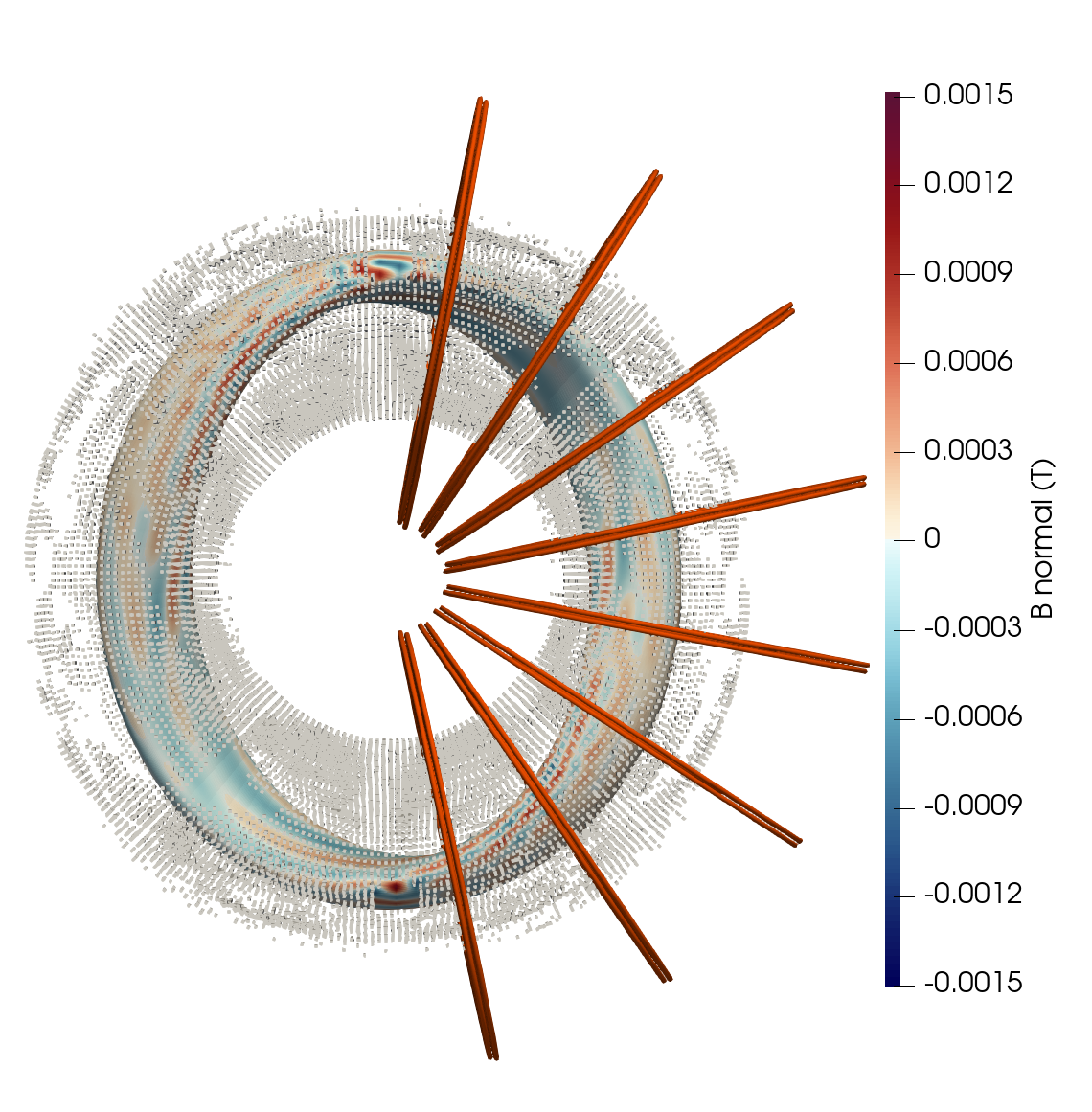}
        \caption{MUSE++}
    \end{subfigure}
    \caption{A comparison among the GPMO solutions of vacuum fields (A), (B), MUSE and MUSE++. The orange planes represent the planar coils. Note that the axis shape of (B) is less elongated.  This potentially contributes to its lower PM count.}
    \label{fig:pm_fancy}
\end{figure}

Finally, Fig.~\ref{fig:pm_fancy} shows a side-by-side comparison of the 4 GPMO solutions. Note that the axis shape of (B) is less elongated than that of MUSE. This better conforms to the shape of the PM holders and leads to a simpler PM array. 
\begin{table}
    \centering
    \begin{tabular}{lc}\toprule
         & Dipole count (per half field period)\\\midrule
         MUSE (GPMO)         & $13861$\\
         MUSE++ (GPMO)& $16801 (+21.2\%)$\\
         (A) (GPMO)& $12841 (-7.3\%)$\\
         (B) (GPMO)& $9121 (-34.2\%)$\\
    \end{tabular}
    \caption{Dipole counts of the GPMO solutions with $f_B=1.29\times10^{-8}\text{T}^2\text{m}^2$.}
    \label{tab:gpmo}
\end{table}

\subsection{Low-force filament coils}
\label{sec: results: force}
To further motivate our QUADCOIL-based QSS method, we now allow the winding surface to move via Algorithm~\ref{alg:ch:formulation:direct} and demonstrate how QSS can be used to improve filament coil performance. The model of this study is the ARIES-CS stellarator~\cite{najmabadi2008aries,aries_cs}, a three-field-period quasi-axisymmetric reactor-scale design with the aspect ratio $A=4.5$, the major radius $R=7.5$m, and the on-axis magnetic field $B_\text{axis} = 5.7$T. One of the most challenging aspects of designing reactor-scale stellarators is the need to keep forces within tolerances during optimization~\cite{hurwitz2025electromagnetic,kaptanoglu2025reactor}. We show in this section that a QSS equilibrium can achieve a $28.3\%$ reduction in RMS coil force and a $30.6\%$ reduction in peak coil force with no reduction in field error and nearly identical physics properties as ARIES-CS.

In this section, we use a force-reducing QUADCOIL proxy:
\begin{equation}\label{eq: quadcoil force}
    \begin{split}
        f_{c, L}(x) =  \int_{S'}dS'\,\|L'_{r, \phi, z}\|_{1},\quad
        x'_* = &\argmin_{x'} \int_{S'}dS'\,\|L'_{r, \phi, z}\|_{1}\\
        &\text{subject to}\\
        &f_B\leq f_{B,\text{target}}=9.62\,\text{T}^2 \text{m}^2,\\
        &|K'|^2\leq K'^2_\text{target},\\
        &K_\theta\leq0
    \end{split}
\end{equation}
Note that in normalized units $\tilde{f}_{B,\text{target}}=9.62\,\,\text{T}^2\text{m}^2 / (B_\text{axis}^2 S) \sim 5\times 10^{-4}$. The L-1 norm in $f_{c, L}$ is chosen as a proxy for the L-2 norm of the Lorentz force, which is quartic in $\Phi$ and does not have the provable near-convexity shown in \cite{lanke_fu_global}.  Unlike in Section \ref{sec: results: PM}, the new equilibrium is constrained to have an identical rotational transform profile as ARIES CS.

\begin{figure}
    \centering
    \includegraphics[width=0.75\linewidth]{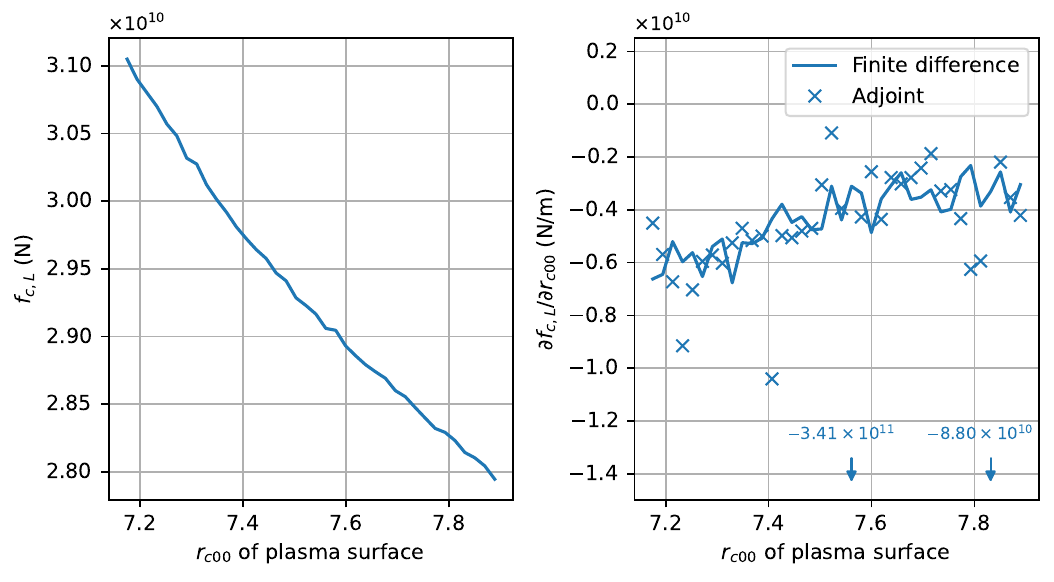}
    \caption{The values (left) and partial derivatives (right) of $f_{c, L}$ w.r.t. the plasma Fourier coefficient $r_{c00}$. The finite difference grid size is $\approx10^{-3}r_{c00}$. }
    \label{fig:ch:single-stage:taylor_force}
\end{figure}

The singular integral in \eqref{eq:ch:formulation:force} is approximately evaluated by removing the quadrature points where the singularity occurs. The target squared flux, $f_{B,\text{target}}$, is empirically chosen based on the optimum of the following QUADCOIL problem on ARIES-CS:
\begin{equation}
    \begin{split}
     f_{B,\text{target}} = 2&\min_{x'}f_B\\
        &\text{subject to}\\
        &|K'|^2\leq K'^2_\text{target},\\
        &K_\theta\leq0.
    \end{split}
\end{equation}
The target squared current density, $K'^2_\text{target}$, is chosen based on the coil current and spacing of ARIES-CS.
\begin{equation}
    K'^2_\text{target} = 2.86\times10^{14}\text{A}^2/\text{m}^{-2} = \left(15.6 \text{MA}/0.77\text{m}\right)^2.
\end{equation}
The winding surface is generated using Alg.~\ref{alg:ch:formulation:direct}. Fig.~\ref{fig:ch:single-stage:taylor_force} shows a Taylor test for $f_{c, L}$. Compared to $f_{c, A}$ and $f_{c, B}$, the values and gradients of $f_{c, L}$ are both non-smooth. This indicates that the augmented Lagrangian solver has reduced accuracy due to the problem's non-convexity. The agreement between the finite difference and adjoint gradients is also poorer than that in Fig.~\ref{fig:ch:single-stage:taylor}. At two points, the adjoint gradient blows up. The potential causes of the inaccuracies are:
\begin{enumerate}
    \item Insufficiently accurate $x_*'$
    \item Incompatibility between JAX autodiff and the singular integral routine \eqref{eq:ch:formulation:force} for certain inputs.
    \item Incompatibility between JAX autodiff and the winding surface generator for certain inputs.
    \item Regions with ill-conditioned $\partial S(x, z')/\partial z'$.
\end{enumerate}
These inaccuracies can cause early terminations of the QSS optimization. We plan to address this issue in future work with further tests, an improved adjoint differentiation scheme, and auto-restart features for the QSS optimization. However, as the following paragraphs show, the QUADCOIL proxy can still lead to significant improvements in filament coil Lorentz force despite the inaccuracies in $x_*'$ and $\nabla f_{c, L}$.

\begin{figure}
    \centering
    \includegraphics[width=1\linewidth]{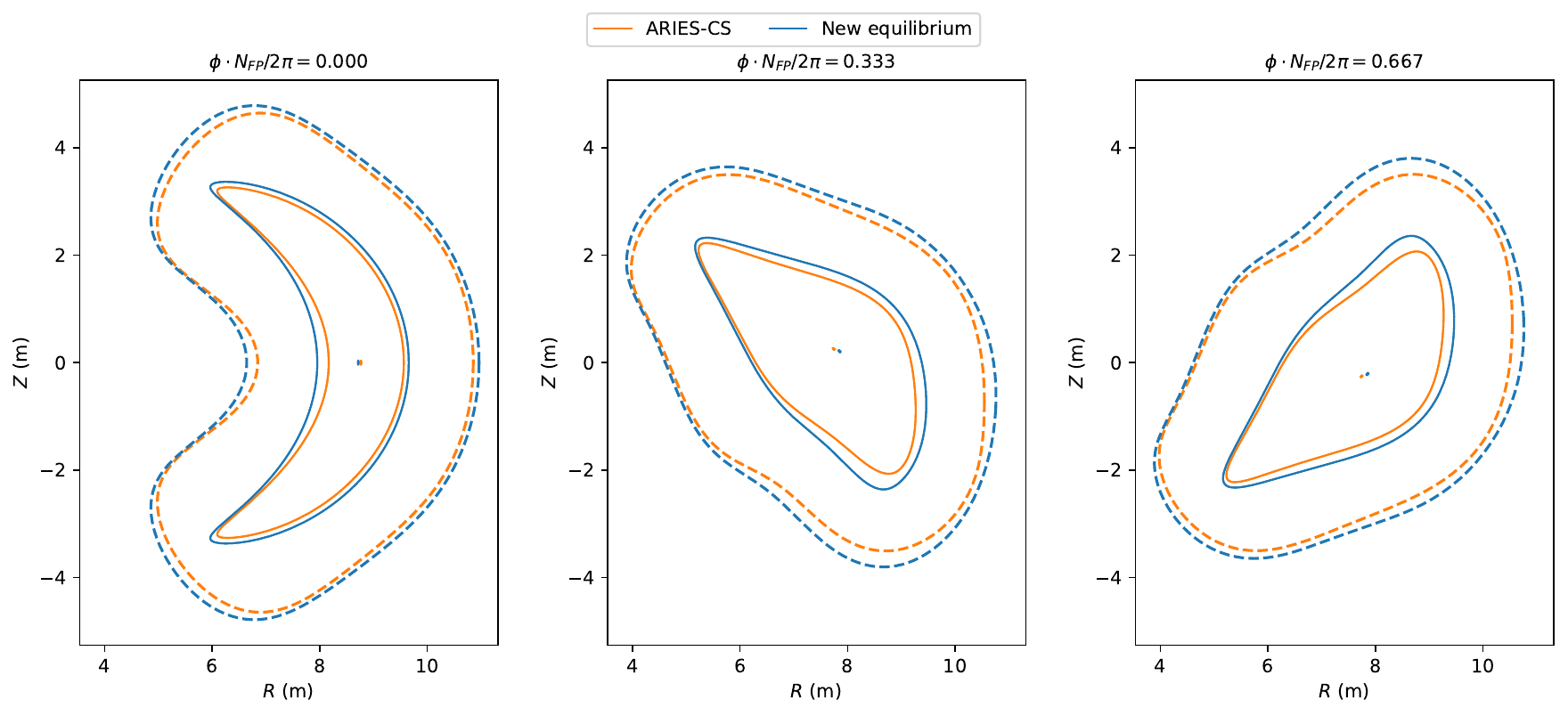}
    \caption{Plasma boundaries (solid), axes, and winding surfaces (dashed) of ARIES-CS and the new equilibrium. The winding surfaces are generated with Alg. \ref{alg:ch:formulation:direct}.}
    \label{fig:ch:single-stage:flux_aries}
\end{figure}

\begin{figure}
    \centering
    \includegraphics[width=0.9\linewidth]{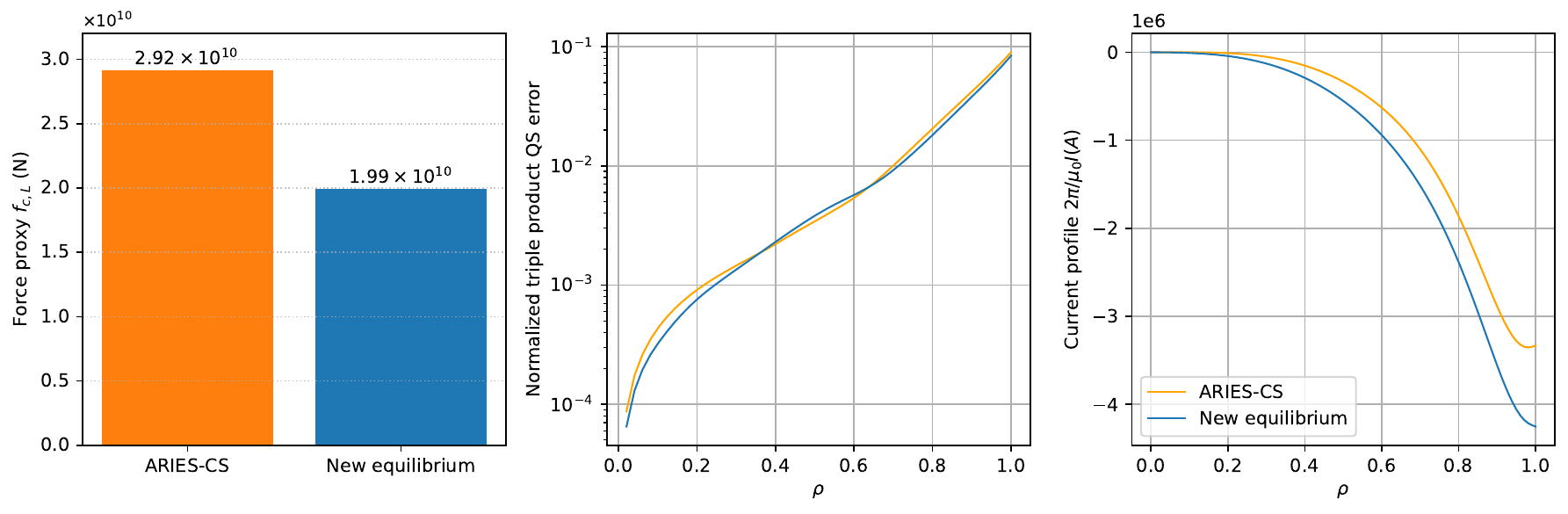}
    \caption{The $f_{c, L}$, QS quality, and current profiles of ARIES-CS and the new equilibrium.}
    \label{fig:ch:single-stage:qs_aries}
\end{figure}

\begin{figure}[htbp]
    \centering
    \begin{subfigure}[b]{0.45\textwidth}
        \includegraphics[width=\textwidth]{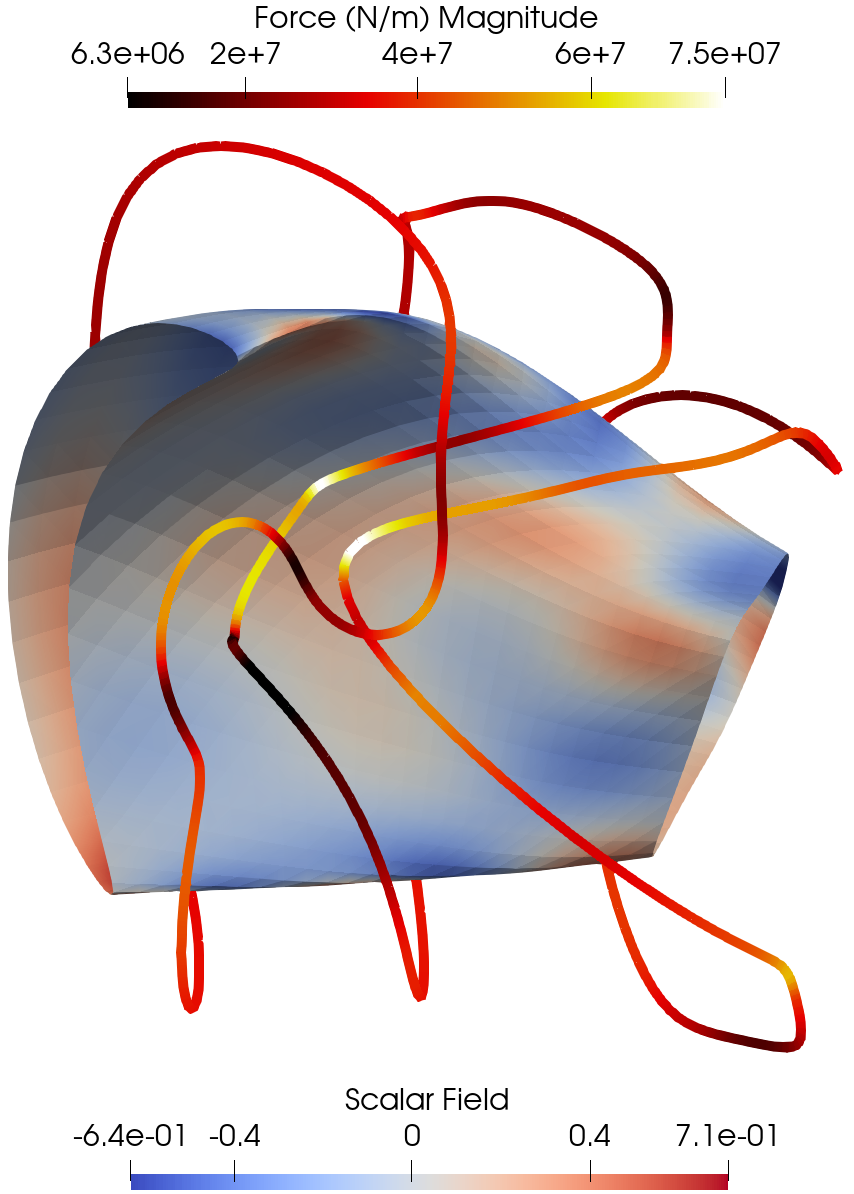}
    \end{subfigure}
    \begin{subfigure}[b]{0.45\textwidth}
        \includegraphics[width=\textwidth]{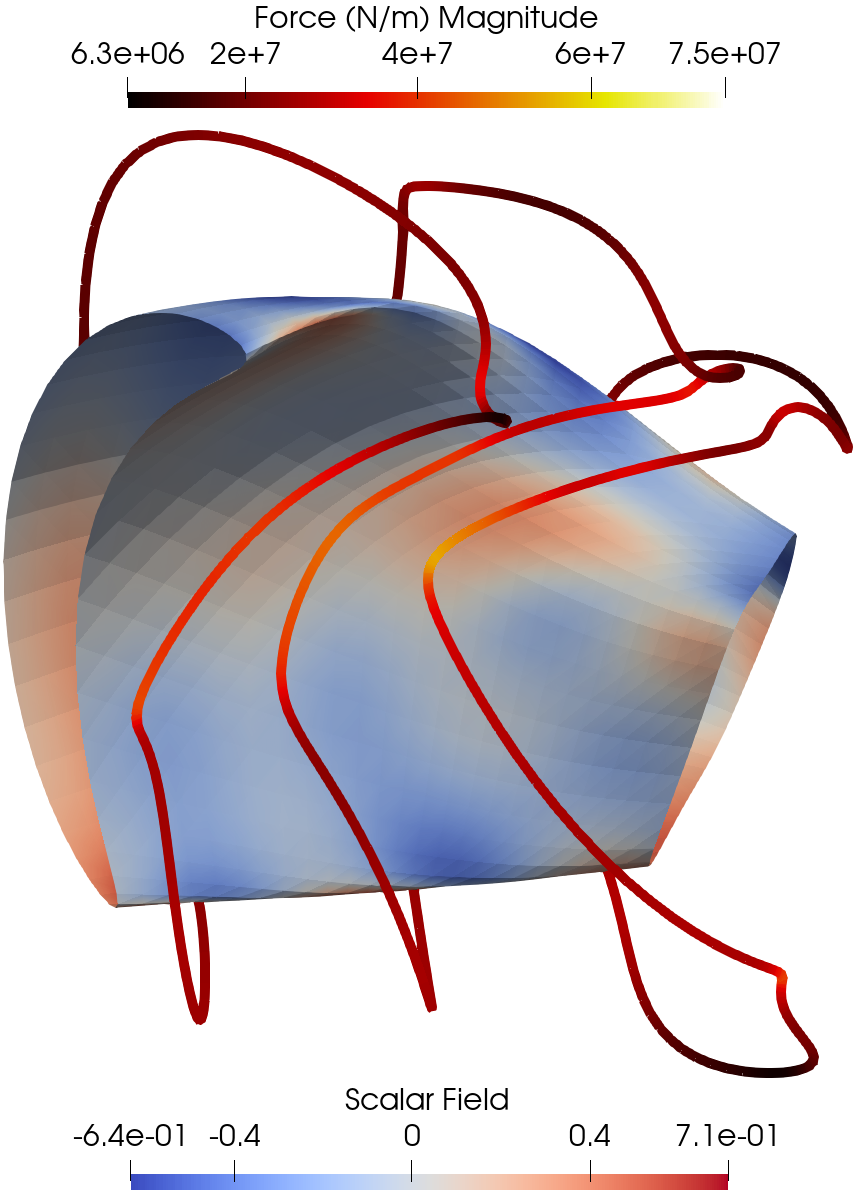}
    \end{subfigure}
    \caption{A comparison between the stage-2 solutions of ARIES-CS (left) and the new equilibrium (right).}
    \label{fig:filament_force}
\end{figure}

Fig.~\ref{fig:ch:single-stage:flux_aries} compares the plasma boundaries of ARIES-CS and the new equilibrium. Since the new equilibrium is constrained to have an identical $\iota$ profile as ARIES-CS, Fig.~\ref{fig:ch:single-stage:qs_aries} compares the $f_{c, L}$, current profiles, and QS quality between ARIES-CS and the new equilibrium. The new equilibrium has almost identical QS quality and current profiles to those of ARIES-CS while achieving a $31.8\%$ improvement in $f_{c, L}$. Fig.~\ref{fig:filament_force} shows the stage-2 solutions and their force distribution. 

The QSS optimization runs in 4.7 hours for 54 iterations on an RTX 4060 GPU. This is much longer than the runtime of the PM QSS optimizations. On average, each iteration takes 313 seconds, and each $f_{c, L}$ evaluation takes 63 seconds. While no published results on force-minimizing single-stage optimization exist, a typical fixed-boundary single-stage optimization in Jorge et al.~\cite{intro_single_vmec_fixed} takes $\sim 20$ minutes on 8 CPUs with 160 GB of total memory~\cite{jorge_rogeriojorgesingle_stage_optimization_2023}. Therefore, the QUADCOIL QSS optimization takes longer to converge but is substantially less resource-demanding. The QSS runtime will likely improve with more accurate adjoint derivatives, faster GPUs, and by using the QUADCOIL solution from the previous QSS step to initialize the augmented Lagrangian solver.

We now validate the effectiveness of the coil force proxy $f_{c, L}$ by removing the winding surface and cold starting a filament stage-2 optimization on ARIES-CS, as well as the new equilibrium found with QSS. The stage-2 objective is:
\begin{equation}
\label{eq:filament_stage_2}
    J \equiv 10^8\max(J_B-5\times10^{-4}, 0)^2+ 5\times10^{-5} J_\text{force}  + 200 J_{l} + 100 J_{\text{cc}} + 100J_{\text{cs}} + 10J_\text{link} + 1000J_\kappa.
\end{equation}
This function contains 5 penalty terms:
\begin{enumerate}
    \item $J_B$, the normalized squared flux:
\begin{equation}
    \label{eq:JB}
    J_B \equiv \frac{1}{2} \frac{\int_S\left(\B_\text{coil} \cdot \n-B_T\right)^2 dS}{\int_S|B|^2 dS}.
\end{equation}
    \item $J_\text{force}$, the L-2 Lorentz force objective \cite{hurwitz2025electromagnetic}:
    \begin{equation}
        J_\text{force} = \frac{1}{2}\sum_i\frac{1}{L_i}\left[\int_{\text{curve}_i} \text{max}(|F'| - F_0 , 0)^2 d\ell_i\right]
    \end{equation}
    where $F'$ is the Lorentz force density and $l'$ is the length of the coil. The target force $F_0$ is set to zero to minimize coil forces as much as possible.
    \item $J_l$, the maximum coil length constraint:
    \begin{equation}
        J_l = \frac{1}{a_{\text{WS}}^2}\sum_{i = 1}^{N_\text{coil}}0.5\max(l'_i - l'_{\text{target}}, 0)^2.
    \end{equation}
    Here, the target maximum coil length is $l_{\text{target}}=31$m. This is shorter than the ARIES-CS value~\cite{aries_cs} for regularization since \eqref{eq:filament_stage_2} contains no curvature or torsion terms.
    \item  $J_{\text{cc}}$, the coil-coil spacing constraint:
    \begin{equation}
            J \equiv \frac{1}{d^{\min}_{\text{cc}}}\sum_{i = 1}^{N_\text{coil}} \sum_{j = 1}^{i-1} \int_{\text{curve}_i} \int_{\text{curve}_j} \max(d^{\min}_{\text{cc}} - \| r'_i - r'_j \|_2, 0)^2 ~dl'_j ~dl'_i,\\
    \end{equation}
Here, the threshold for the minimum coil-coil distance is $d^{\min}_{\text{cc}}=0.77\text{m}$, based on the ARIES-CS value.
    \item $J_{\text{cs}}$, the coil-plasma spacing constraint:
    \begin{equation}
        J_{\text{cs}} \equiv \frac{1}{a_\text{winding}} \sum_{i = 1}^{N_\text{coil}} \int_{\text{curve}_i} \int_{\text{surface}} \max(d^{\min}_{\text{cs}} - \| r'_i - s \|_2, 0)^2 ~dl'_i ~ds.
    \end{equation}
Here, $a_\text{winding}$ is the minor radius of the winding surface. The threshold for the minimum coil-plasma distance is $d_{cs}^\text{min}=1.3\text{m}$, based on the ARIES-CS value.
    \item $J_{\text{link}}$, the linking number constraint. Each pair of curves contributes:
    \begin{equation}
        \text{Link}(c_1, c_2) = \frac{1}{4\pi} \left| \oint_{c_1}\oint_{c_2}\frac{\rr'_1 - \rr'_2}{|\rr'_1 - \rr'_2|^3} (d\rr'_1 \times d\rr'_2) \right|.
    \end{equation}
    \item $J_\kappa$, the L-2 curvature objective:
    \begin{equation}
        J_\kappa = \frac{1}{2}\sum_{i = 1}^{N_\text{coil}} \int_{\text{curve}_i} \text{max}(\kappa - \kappa_0, 0)^2 ~dl
    \end{equation}
Here, the target curvature $\kappa_0$ is set to $1/(30\text{cm})$, based on the radius of curvature in the original ARIES-CS coil design.
\end{enumerate}
The objective function \eqref{eq:filament_stage_2} will reduce $J_B$ to an empirically chosen target, $5\times10^{-4}$, but not further. Then, it will attempt to reduce the mean squared coil forces. 

\begin{figure}
    \centering
    \includegraphics[width=0.75\linewidth]{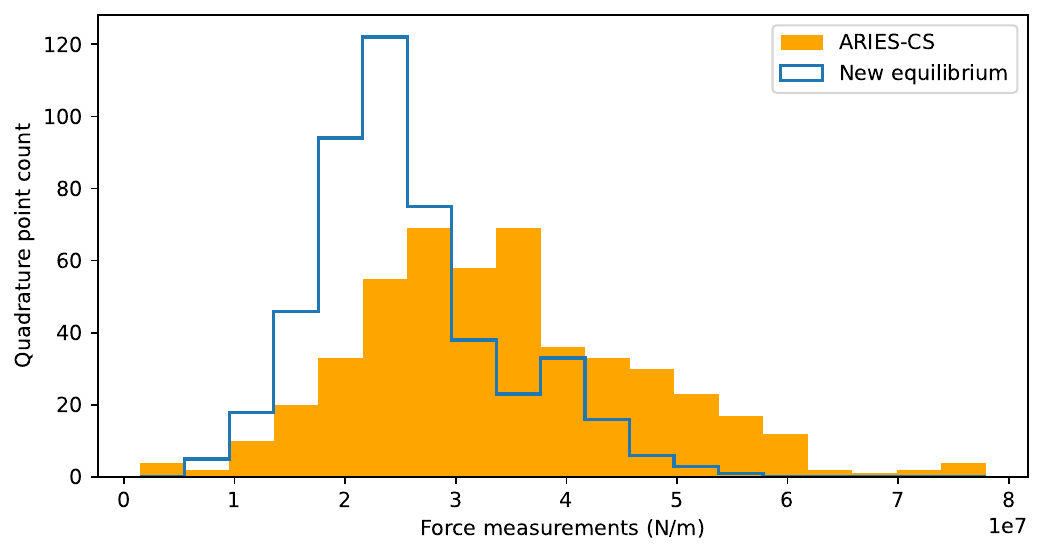}
    \caption{The histogram of the Lorentz force strengths in the stage-2 solutions.}
    \label{fig:force:hist}
\end{figure}

Tab.~\ref{tab:force} shows the results of the filament optimization. Compared to ARIES-CS, the filament coil set for the new equilibrium achieves a $28.3\%$ reduction in RMS force and a $30.6\%$ reduction in peak force with the same squared flux. Fig.~\ref{fig:filament_force} shows the force distributions on the stage-2 solutions. Fig.~\ref{fig:force:hist} shows a histogram of the measured force strengths at the quadrature points of the stage-2 solutions. Both plots show that the new equilibrium's coil set has a lower Lorentz force overall. Note that the winding surface geometry and surface current configurations are entirely discarded after the QSS optimization. Therefore, the key takeaway is that the compatibility of the new plasma surface with low-force coils remains, even though the filamentary coils are optimized very differently from the original winding surface solution.

\begin{table}
    \centering
    \begin{tabular}{ccc}\toprule
         &  ARIES-CS& The new equilibrium\\
         \midrule
 Normalized squared flux& $5.11\times10^{-4}$&$5.06\times10^{-4}$\\
         RMS force $(\text{MN}/\text{m})$&  $35.4$& $25.4 (-28.3\%)$\\
         Peak force $(\text{MN}/\text{m})$&  $77.9$& $54.0 (-30.6\%)$\\ 
 Max length $(\text{m})$& $31.0$&$31.0$\\
 Min coil-coil distance $(\text{m})$& $0.767$&$0.769$\\
 Min coil-plasma distance $(\text{m})$& $1.22$&$1.24$\\ 
 Min radius of curvature $(\text{m})$& $0.300$&$0.300$\\ \bottomrule
    \end{tabular}
    \caption{Stage-2 optimization results of ARIES-CS and the new equilibrium.}
    \label{tab:force}
\end{table}

\section{Conclusion and outlooks}\label{sec: conclusion}

In this paper, we have developed QUADCOIL into a flexible, differentiable coil complexity proxy for equilibrium optimization. We have also presented two numerical studies that demonstrate the effectiveness of the QUADCOIL proxy in simplifying both PM/dipole arrays and filament coils. We believe the results of QUADCOIL QSS can be further refined with single-stage optimization to design high-performance stellarators with low engineering costs. Our implementation is robust for convex QUADCOIL problems. For non-convex QUADCOIL subproblems, our implementation of the gradient of the proxy is not yet robust, but the QSS optimization still leads to significant improvements. Notably, the current implementation of the QUADCOIL proxy cold-starts the augmented Lagrangian solver in every evaluation. Future implementations will use the QUADCOIL solution from the previous QSS step to initialize the augmented Lagrangian solver to improve speed and accuracy.

There are three future paths toward improving the combined coil-plasma optimization with QUADCOIL. The first path for improvement is to enhance the accuracy of the QCQP solver and improve the robustness of QUADCOIL in auto-differentiation. The second path is to develop custom stationarity conditions with robust constraint handling. The third path is to perform single-stage optimization by simultaneously solving for the QUADCOIL and plasma degrees of freedom with a single term that couples them, as in the Jorge et al. formulation~\cite{intro_single_vmec_fixed3}:
\begin{align}\label{eq:quasi-single example2}
        &\min_{x, x'} \;
            \omega_T\hat f_\text{T}(x) 
            +\omega_{\iota}f_{\iota}(x)
            +\omega_c\Phi'_\text{max}(x'),\\ \notag 
        &\text{subject to:}\quad g_p(x) \leq 0, \quad f_B(x, x')\leq f_{0}
\end{align}
While this approach requires more degrees of freedom than QSS, the minimization over $x'$ is convex if we restrict the problem to convex objectives with respect to $x'$. This approach no longer requires a QUADCOIL solve per iteration. This can help improve iteration speed. However, the speed-up may also become bottlenecked by the equilibrium solver and optimizer overheads. The comparative advantages of QSS and single-stage optimization using the winding surface model require further investigation.

\section*{Acknowledgments}
This work is supported by the Department of Energy under the HifiStell SciDAC grant (DE-SC0024548) and the Simons Foundation under award 560651. We are grateful to Xu Chu for providing the MUSE configuration, Caoxiang Zhu and Guodong Yu for providing the MUSE++ configuration, and other members of the Simons Collaboration on Hidden Symmetries and Fusion Energy. 

\section*{Data availability statement}
The data that support the findings of this study is openly available~\cite{Lanke2025-re}.

\appendix
\numberwithin{equation}{section}
\renewcommand{\theequation}{\thesection\arabic{equation}}
\section{Penalty and barrier method formulations}
\label{sec:appendix_penalty_barrier}
There are many ways to solve nonconvex constrained optimization problems, and there are advantages and disadvantages for each algorithm.

For completeness, we now discuss possible penalty or barrier formulations for the QUADCOIL solver. Both methods still convert the constrained optimization problem to an unconstrained problem. A penalty method performs this conversion by solving:
\begin{equation}
    \begin{split}
        \min_{x'}&\left[f_c(x') + w_\text{ineq}\sum_i\left|\min\{0, (g_c)_i(x')\}\right|^2 + w_\text{eq}\left\|h_c(x')\right\|^2_2\right],
        \\
        S_\text{penalty} & \equiv\partial f_\text{penalty}/\partial x' = 0,
    \end{split}
\end{equation}
where $w_\text{ineq}$ and $w_\text{eq}$ are weight factors. A barrier method converts a constrained problem into an unconstrained problem by replacing inequality constraints with barrier functions that grow to infinity when the constraints are violated. One commonly used example is the log barrier:
\begin{equation}
\begin{split}
    \min_{x', t}&\left[ct - \log(t-f_c(x'))-\sum_i\log(-(g_c)_i(x'))\right],\\
    S_\text{barrier} & \equiv\partial f_\text{barrier}/\partial x'  + f_\text{barrier}/\partial t  = 0,
\end{split}
\end{equation}
where $c$ is a steepness factor, $t$ is a slack variable, and the equality constraints $h_c$ are removed for the simplicity of discussion. A special advantage of the log barrier method is that for convex QCQP, $S_\text{barrier}$ is provably non-singular everywhere if it is non-singular for one $(x', t)$~\cite{nesterov_interior-point_1994}. The penalty and barrier methods share the similarity that $w_\text{ineq}, w_\text{eq}$, and $c$ must be arbitrarily large to tightly satisfy the inequality constraints. Because of this, $S_\text{penalty}$ and $S_\text{barrier}$ are often numerically ill-conditioned at the optimum, even when $S_\text{barrier}$ is provably non-singular~\cite{book_nocedal_numerical_2006}. This makes adjoint differentiation difficult with both methods. In contrast, our approach does not suffer from this issue because $\mu$ and $\lambda$ do not need to be arbitrarily large in the augmented Lagrangian method. Of course, one can relax the permitted amount of constraint violation to improve the conditioning of $S_\text{penalty}$ and $S_\text{barrier}$. In practice, a constraint violation of $\sim1\%$ is often sufficiently low. However, this may require manual tuning of the penalty weight for each combination of constraints and parameters. The trade-off between QUADCOIL constraints and objectives is also not well understood. It is possible that a small increase in constraint violation can substantially alter the resulting coil configuration. Discussions on this is beyond the scope of the present paper. 

\section{Limitations when the number of constraints is very large}
\label{sec:appendix_limitations}
During our numerical study, we found it challenging to reliably perform adjoint differentiation on problems with large $\numineq$. These challenges reflect the inherent limitations of the augmented Lagrangian method. QUADCOIL can accurately differentiate \eqref{eq: quadcoil A}, a single-constraint problem, but not \eqref{eq: quadcoil B}, a problem with $\mathcal{O}(\numgrid')$ constraints. This is likely due to difficulties in tracking active constraints. In this section, we will present evidence that supports our hypothesis and illustrate the challenges in differentiating a constrained optimization problem.

To inspect how adjoint derivatives may fail, we use a slack variable $s$ to convert \eqref{eq: quadcoil B} into the $C^2$ problem that QUADCOIL solves: 
\begin{equation}
\label{eq:ch:single-stage:thickness2}
    \min_{x', s} \,\, s,\quad
    \text{subject to:}\quad f_B(x')\leq f_{0},\quad
    \Phi'(\zeta'_{ij}, \theta'_{ij})(x')-s\leq0,\quad
    -\Phi'(\zeta'_{ij}, \theta'_{ij})(x')-s\leq0.
\end{equation}
Here, $\Phi'$ is the dipole density distribution on the winding surface, and $s$ is a slack variable. This is an optimization problem with a linear objective, one quadratic inequality constraint, and $O(\numgrid')$ inequality constraints. 

\begin{figure}
    \centering
    \includegraphics[width=0.5\linewidth]{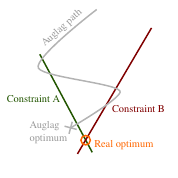}
    \caption{The augmented Lagrangian method does not exactly track the active constraint set. In this figure, both A and B should be exactly satisfied and active at the optimum (marked with $\circ$), but an augmented Lagrangian solver may converge to a point where only constraint A is active (marked with $\times$). The active set at the numerical solution can vary depending on the initial state and optimization path (shown in gray). At $\times$, the multiplier corresponding to constraint B is zero. Because of this, the adjoint derivative obtained at $\times$ will also be inaccurate.}
    \label{fig:ch:single-stage:auglag}
\end{figure}

\begin{figure}
    \centering
    \includegraphics[width=0.95\linewidth]{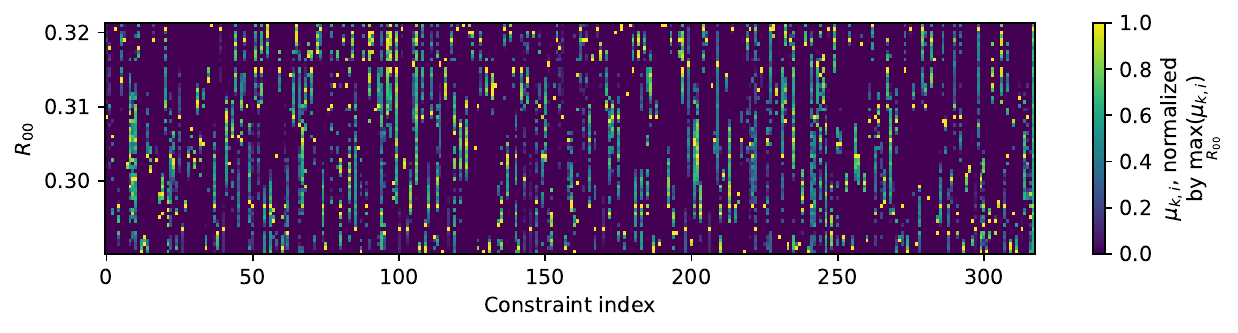}
    \caption{The multiplier $\mu_{k, i}$ values across all $r_{c00}$. Each column represents the change of one $\mu_{k}$ component with $r_{c00}$. The colors in this plot is normalized by the maximum value of each component across all $r_{c00}$. }
    \label{fig:ch:single-stage:activity}
\end{figure}

We then write down the stationarity condition by substituting in the definition of $L_k$ from \eqref{eq:ch:numerical:auglag}:
\begin{align}
\notag
        S = 0 = \,
         \partial_{x'}f_c(x')+&\lambda_k^\top \partial_{x'}h_c(x')
        +\textcolor{black}{\mu_k^\top \partial_{x'}g^+_c(x', \mu_k, c_k)} 
         + c_k  \partial_{x'}h_c(x')^\top h_c(x')
         +c_k\partial_{x'}g^{+}_c(x', \mu_k, c_k)^\top g^{+}_c(x', \mu_k, c_k),\\
         \text{where } (g^+_c)_j\equiv&\max\{(g_c)_j(x'), -(\mu_k)_j/c_k\}.
         \label{eq:ch:single-stage:sta2}
\end{align}
This adjoint derivative can become inaccurate when $\mu_{k, i}=0$ for a constraint $(g_c)_i$ that should be exactly satisfied and active. When this happens, the third term above becomes zero, and some gradient information about $(g_c)_i$ becomes lost. However, this is challenging to prevent in an augmented Lagrangian method. When an augmented Lagrangian solve converges, constraints that should be exactly satisfied can often become inactive. (Fig.~\ref{fig:ch:single-stage:auglag}) This can occur even when the solver converges with sufficient tolerance. When important constraints become inactive, information from these constraints will be lost from $S$, causing incorrect gradients or singular $\partial_{z'} S$. One may suggest switching to a barrier or penalty method for solving \eqref{eq:problem}. Theoretically, barrier/penalty methods do not suffer from this issue because all constraints are active at all times. However, in practice, the Hessian of these methods can often become near singular at the optimum, which makes \eqref{eq:ift} numerically challenging to evaluate.
To confirm this hypothesis, Fig.~\ref{fig:ch:single-stage:activity} shows the final values of $\mu_{k}$ \eqref{eq:ch:single-stage:thickness2} across all $r_{c00}$. Each column in Fig.~\ref{fig:ch:single-stage:activity} corresponds to the change of a different component in $\mu_k$ with $r_{c00}$. For simplicity, we have omitted $\sim1700$ constraints that stayed inactive for all $r_{c00}$. Inspecting the columns in this plot, we find that a component $\mu_{k, i}$ can become zero/non-zero sporadically with small changes in $r_{c00}$. This means the corresponding constraint, $(g_c)_i$, can also become sporadically inactive/active for small changes in $r_{c00}$. Intuitively, the activation/deactivation of $(g_c)_i$ should be smooth for smooth changes in $r_{c00}$. While not conclusive, this supports our hypothesis that the adjoint derivatives are inaccurate, at least partially, due to the imperfect tracking of inactive/active inequality constraints. 

\section{QSS with MUSE++ initial condition}
\label{sec:appendix_musepp}
We performed an additional investigation, as in Section~\ref{sec: results: PM}, in order to investigate if we could achieve quasi-axisymmetry performance comparable with the MUSE++ design, but still reducing the dipole density on the winding surface.
When using MUSE++ as the initial condition, the QSS objective is:
\begin{equation}
\begin{gathered}
        \min_x \left\{
            \omega_\text{V}[V(x) - V_\text{MUSE++}]^2 
            + \omega_\iota[\iota_{\text{axis}}(x) - \iota_{\text{axis, MUSE++}}]^2 
            +\omega_\iota[\iota_{\text{edge}}(x) - \iota_{\text{edge, MUSE++}}]^2 
            +\omega_\text{T}\hat f_\text{T}(x) 
            +\omega_cf_c(x)\right\},\\
            \text{subject to } r_{c00} = (r_{c00})_\text{MUSE++} ,\psi_\text{LCFS} = \left(\psi_\text{LCFS}\right)_\text{MUSE++} .
\end{gathered}
\end{equation}
The QUADCOIL proxy $f_{c, A}$ and $f_{c, B}$ remain unchanged. We refer to the new vacuum fields as (A2) and (B2).

\begin{figure}
    \centering
    \includegraphics[width=1\linewidth]{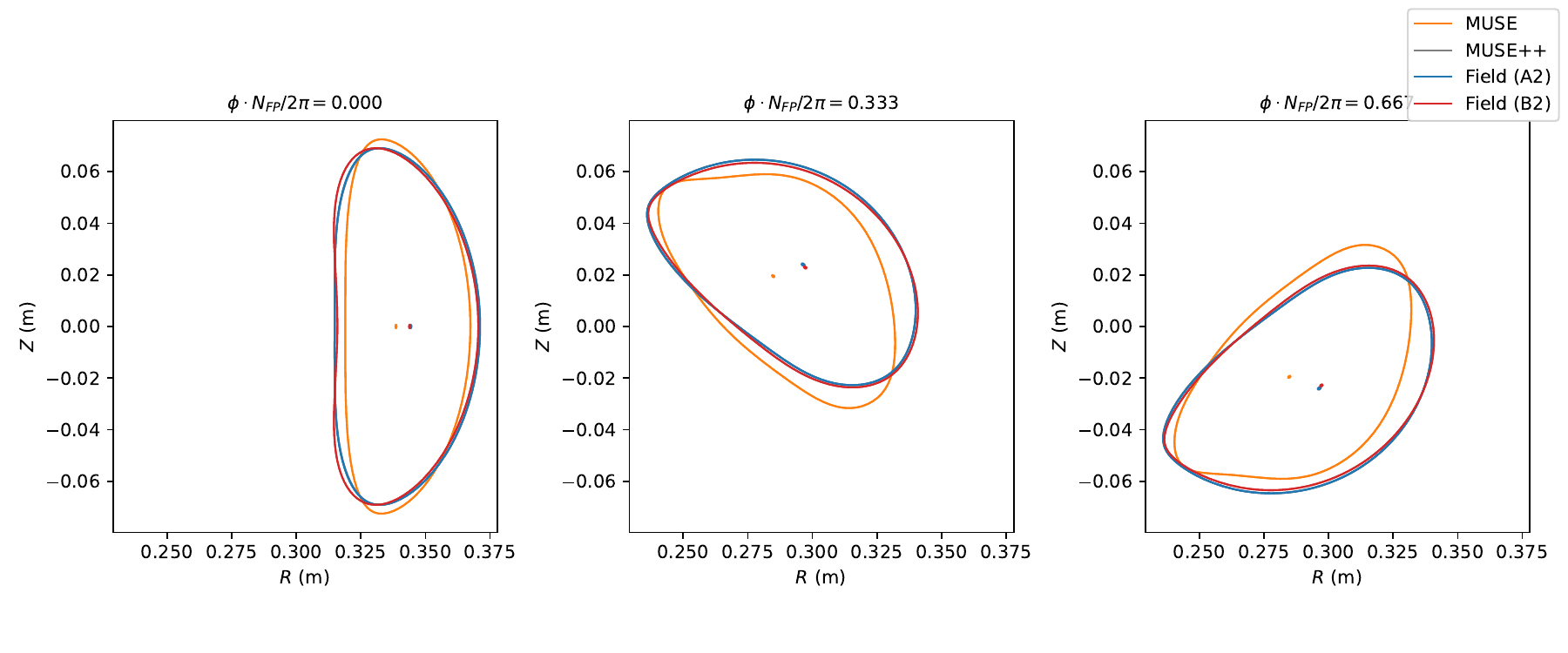}
    \caption{Plasma boundaries of MUSE, MUSE++, (A2) and (B2). Note that both (A2) and (B2) converged close to the initial condition.}
    \label{fig:ch:single-stage:flux2}
\end{figure}

\begin{figure}
    \centering
    \includegraphics[width=0.96\linewidth]{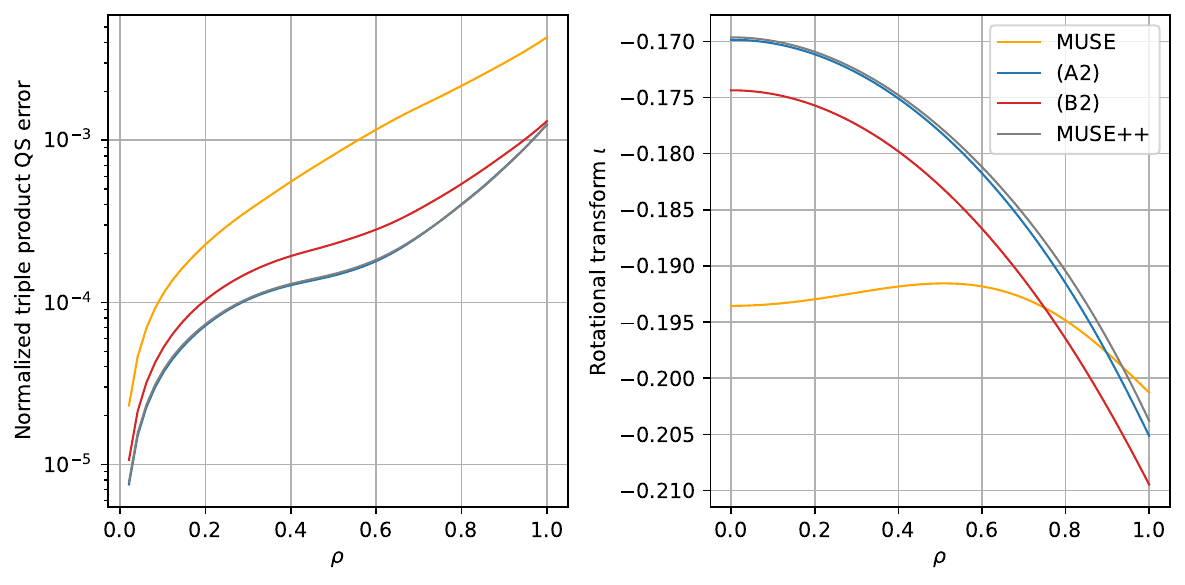}

    \caption{The QS quality values (left) and rotational transform (right) of MUSE, MUSE++ and the new vacuum fields. }
    \label{fig:ch:single-stage:qs2}
\end{figure}

\begin{figure}
    \centering
    \includegraphics[width=0.7\linewidth]{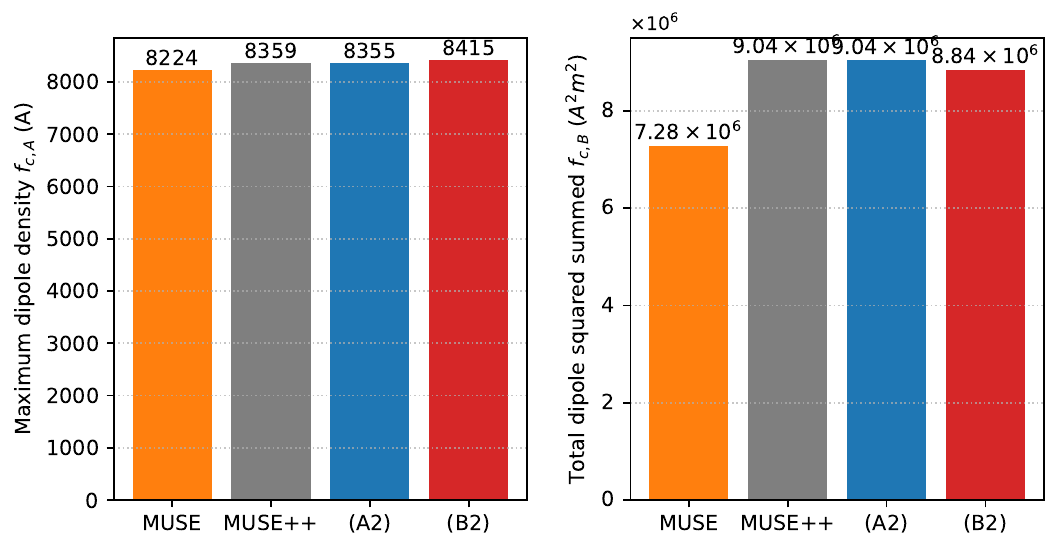}
    \vspace{1em}
    \caption{Comparison of the dipole thickness and count among MUSE, MUSE++, and the two new vacuum fields. Note that (A2) and (B2) only marginally improved QUADCOIL proxy compared to MUSE++}
    \label{fig:bar2}
\end{figure}

Fig.~\ref{fig:ch:single-stage:flux2} compares the outer flux surfaces of MUSE, MUSE++, (A2), and (B2). Fig.~\ref{fig:ch:single-stage:qs2} compares the rotational transform and QS quality among the four vacuum fields. Figure~\ref{fig:bar2} and Table \ref{tab:proxy2} compares the values of $f_{c, A}$ and $f_{c, B}$ among the four vacuum fields. All three plots indicate that the QSS optimization converged close to MUSE++ in both cases, making only marginal improvements in $f_{c, A}$ and $f_{c, B}$. This suggests that the MUSE++ vacuum field may be a narrow local minimum in the QS quality landscape. It appears that in order to achieve the very high QS exhibited by MUSE++, there remains very little slack for improving other metrics such as the dipole density. 

\begin{table}
    \centering
    \begin{tabular}{ccccc}\toprule
         &  MUSE&  MUSE++&  Vacuum field (A2)& Vacuum field (B2)\\\midrule
         \makecell{Maximum dipole\\density $f_{c,A}$ (A)}&  $8224$&  $8359$&  $8355$& $8415$\\
         \makecell{Dipole count\\$f_{c,B}$(A$\text{m}^2$)}&  $7.28\times10^6$&  $9.04\times10^6$&  $9.04\times10^6$& $8.84\times10^6$\\ \bottomrule
    \end{tabular}
    \caption{Comparison of the QUADCOIL proxy of MUSE, (A) and (B) with the MUSE++ values.}
    \label{tab:proxy2}
\end{table}

\bibliographystyle{unsrt}  
\bibliography{references}

@ARTICLE{lbfgs,
  title     = "On the limited memory {BFGS} method for large scale optimization",
  author    = "Liu, Dong C and Nocedal, Jorge",
  journal   = "Math. Program.",
  publisher = "Springer Science and Business Media LLC",
  volume    =  45,
  number    = "1-3",
  pages     = "503--528",
  month     =  aug,
  year      =  1989,
  language  = "en"
}

@book{book_nocedal_numerical_2006,
	address = {New York, NY},
	edition = {Second edition},
	series = {Springer series in operations research and financial engineering},
	title = {Numerical optimization},
	isbn = {978-0-387-30303-1 978-0-387-40065-5},
	language = {en},
	publisher = {Springer},
	author = {Nocedal, Jorge and Wright, Stephen J.},
	year = {2006},
}

@article{guinchard2024including,
  title={Including the vacuum field energy in stellarator coil design},
  author={Guinchard, S and Hudson, SR and Paul, EJ},
  journal={arXiv e-prints},
  pages={arXiv--2409},
  year={2024}
}

@article{ulrich2025permanent,
  title={Permanent magnet optimization of stellarators with coupling from finite permeability and demagnetization effects},
  author={Ulrich, Armin and Haberle, Mason and Kaptanoglu, Alan A},
  journal={arXiv preprint arXiv:2512.14997},
  year={2025}
}

@article{kruger2025coil,
  title={Coil optimization methods for a planar coil stellarator},
  author={Kruger, TG and Martin, MF and Gates, DA and Thea Energy Team},
  journal={Nuclear Fusion},
  volume={65},
  number={2},
  pages={026051},
  year={2025},
  publisher={IOP Publishing}
}

@article{swanson2025scoping,
  title={{The scoping, design, and plasma physics optimization of the Eos neutron source stellarator}},
  author={Swanson, CPS and Gates, DA and Kumar, STA and Martin, MF and Kruger, TG and Dudt, DW and Bonofiglo, PJ and Thea Energy team},
  journal={Nuclear Fusion},
  volume={65},
  number={2},
  pages={026053},
  year={2025},
  publisher={IOP Publishing}
}

@inproceedings{zarnstorff2023stellarex,
  title={{The Stellarex Path to a Stellarator Fusion Pilot Plant}},
  author={Zarnstorff, Michael and Bhattacharjee, A and Carty, R and Team, Stellarex},
  booktitle={APS Division of Plasma Physics Meeting Abstracts},
  volume={2023},
  pages={CM06--009},
  year={2023}
}

@ARTICLE{intro_w7x_delay,
  title     = "Fabrication of the superconducting coils for Wendelstein {7-X}",
  author    = "Risse, Konrad and Rummel, Th and Wegener, L and Holzth{\"u}m, R
               and Jaksic, N and Kerl, F and Sapper, J",
  journal   = "Fusion Eng. Des.",
  publisher = "Elsevier BV",
  volume    = "66-68",
  pages     = "965--969",
  month     =  sep,
  year      =  2003,
  language  = "en"
}

@article{intro_pressure_limit,
    author = {A. Weller and S. Sakakibara and K. Y. Watanabe and K. Toi and J. Geiger and M. C. Zarnstorff and S. R. Hudson and A. Reiman and A. Werner and C. Nührenberg and S. Ohdachi and Y. Suzuki and H. Yamada and and and},
    title = {Significance of MHD Effects in Stellarator Confinement},
    journal = {Fusion Science and Technology},
    volume = {50},
    number = {2},
    pages = {158--170},
    year = {2006},
    publisher = {American Nuclear Society},
    doi = {10.13182/FST06-A1231},
    URL = {https://doi.org/10.13182/FST06-A1231},
    eprint = {https://doi.org/10.13182/FST06-A1231}
}

@article{intro_JET_disruption,
	title        = {Overview of Disruptions with JET-ILW},
	author       = {Gerasimov, S.N. and Abreu, P. and Artaserse, G. and Baruzzo, M. and Buratti, P. and Carvalho, I.S. and Coffey, I.H. and De La Luna, E. and Hender, T.C. and Henriques, R.B. and others},
	year         = 2020,
	month        = jun,
	journal      = {Nuclear Fusion},
	volume       = 60,
	number       = 6,
	pages        = {066028},
	doi          = {10.1088/1741-4326/ab87b0},
	issn         = {0029-5515, 1741-4326},
	url          = {https://iopscience.iop.org/article/10.1088/1741-4326/ab87b0},
	urldate      = {2021-02-16}
}

@article{intro_Menard_2011,
    doi = {10.1088/0029-5515/51/10/103014},
    url = {https://dx.doi.org/10.1088/0029-5515/51/10/103014},
    year = {2011},
    month = {aug},
    publisher = {},
    volume = {51},
    number = {10},
    pages = {103014},
    author = {Menard, J.E. and Bromberg, L. and Brown, T. and Burgess, T. and Dix, D. and El-Guebaly, L. and Gerrity, T. and Goldston, R.J. and Hawryluk, R.J. and Kastner, R. and Kessel, C. and Malang, S. and Minervini, J. and Neilson, G.H. and Neumeyer, C.L. and Prager, S. and Sawan, M. and Sheffield, J. and Sternlieb, A. and Waganer, L. and Whyte, D. and Zarnstorff, M.},
    title = {Prospects for pilot plants based on the tokamak, spherical tokamak and stellarator},
    journal = {Nuclear Fusion}
}

@article{intro_muse,
	title = {Design and construction of the {MUSE} permanent magnet stellarator},
	volume = {89},
	issn = {0022-3778, 1469-7807},
	url = {https://www.cambridge.org/core/journals/journal-of-plasma-physics/article/design-and-construction-of-the-muse-permanent-magnet-stellarator/3028EBD0A22067E0ECE1EF22D9725BFF},
	doi = {10.1017/S0022377823000880},
	language = {en},
	number = {5},
	urldate = {2025-06-05},
	journal = {Journal of Plasma Physics},
	author = {Qian, T. M. and Chu, X. and Pagano, C. and Patch, D. and Zarnstorff, M. C. and Berlinger, B. and Bishop, D. and Chambliss, A. and Haque, M. and Seidita, D. and Zhu, C.},
	month = oct,
	year = {2023},
	pages = {955890502},
}

@article{intro_dipole2,
	title = {Stellarator fusion systems enabled by arrays of planar coils},
	volume = {65},
	issn = {0029-5515, 1741-4326},
	url = {https://iopscience.iop.org/article/10.1088/1741-4326/ada56c},
	doi = {10.1088/1741-4326/ada56c},
	number = {2},
	urldate = {2025-06-05},
	journal = {Nuclear Fusion},
	author = {Gates, D.A. and Aslam, S. and Berzin, B. and Bonofiglo, P. and Cote, A. and Dudt, D.W. and Flom, E. and Fort, D. and Koen, A. and Kruger, T.G. and Kumar, S.T.A. and Martin, M.F. and Ottaviano, A. and Pasmann, S. and Romano, P.K. and Swanson, C.P.S. and Tang, L. and Winkler, E. and Wu, R.},
	month = feb,
	year = {2025},
	pages = {026052},
}

@ARTICLE{intro_pm_opt,
  title     = "Topology optimization of permanent magnets for stellarators",
  author    = "Zhu, Caoxiang and Hammond, Kenneth and Brown, Thomas and Gates,
               David and Zarnstorff, Michael and Corrigan, Keith and Sibilia,
               Marc and Feibush, Eliot",
  journal   = "Nucl. Fusion",
  publisher = "IOP Publishing",
  volume    =  60,
  number    =  10,
  pages     = "106002",
  month     =  oct,
  year      =  2020,
  copyright = "http://iopscience.iop.org/page/copyright"
}

@ARTICLE{lanke_fu_global,
  title     = "Global stellarator coil optimization with quadratic constraints
               and objectives",
  author    = "Fu, Lanke and Paul, Elizabeth J and Kaptanoglu, Alan A and
               Bhattacharjee, Amitava",
  journal   = "Nucl. Fusion",
  publisher = "IOP Publishing",
  volume    =  65,
  number    =  2,
  pages     = "026045",
  month     =  feb,
  year      =  2025,
  copyright = "https://creativecommons.org/licenses/by/4.0/"
}

@ARTICLE{intro_single_vmec_fixed,
  title     = "Simplified and flexible coils for stellarators using
               single-stage optimization",
  author    = "Jorge, R and Giuliani, A and Loizu, J",
  journal   = "Phys. Plasmas",
  publisher = "AIP Publishing",
  volume    =  31,
  number    =  11,
  month     =  nov,
  year      =  2024,
  copyright = "https://creativecommons.org/licenses/by/4.0/",
  language  = "en"
}

@techreport{pomphrey2000compact,
  title={Compact stellarator coils},
  author={Pomphrey, N and Berry, L and Boozer, A and Brooks, A and Hatcher, R},
  year={2000},
  institution={Princeton Plasma Physics Lab.(PPPL), Princeton, NJ (United States)}
}

@ARTICLE{intro_single_vmec_fixed3,
  title     = "Single-stage stellarator optimization: combining coils with
               fixed boundary equilibria",
  author    = "Jorge, R and Goodman, A and Landreman, M and Rodrigues, J and
               Wechsung, F",
  journal   = "Plasma Phys. Control. Fusion",
  publisher = "IOP Publishing",
  volume    =  65,
  number    =  7,
  pages     = "074003",
  month     =  jul,
  year      =  2023,
  copyright = "http://creativecommons.org/licenses/by/4.0"
}

@misc{jorge_rogeriojorgesingle_stage_optimization_2023,
	title = {rogeriojorge/single\_stage\_optimization: {Tag} for the zenodo release},
	url = {https://doi.org/10.5281/zenodo.7655077},
	doi = {10.5281/zenodo.7655077},
	publisher = {Zenodo},
	author = {Jorge, Rogerio and Goodman, Alan},
	month = feb,
	year = {2023},
}

@ARTICLE{intro_single_spec_free,
  title     = "Free-boundary {MRxMHD} equilibrium calculations using the
               stepped-pressure equilibrium code",
  author    = "Hudson, S R and Loizu, J and Zhu, C and Qu, Z S and
               N{\"u}hrenberg, C and Lazerson, S and Smiet, C B and Hole, M J",
  journal   = "Plasma Phys. Control. Fusion",
  publisher = "IOP Publishing",
  volume    =  62,
  number    =  8,
  pages     = "084002",
  month     =  aug,
  year      =  2020,
  copyright = "http://iopscience.iop.org/page/copyright"
}

@article{hammond2024improved,
  title={Improved stellarator permanent magnet designs through combined discrete and continuous optimizations},
  author={Hammond, KC and Kaptanoglu, AA},
  journal={Computer Physics Communications},
  pages={109127},
  year={2024},
  publisher={Elsevier}
}

@ARTICLE{intro_single_overview,
  title     = "Combined plasma--coil optimization algorithms",
  author    = "Henneberg, S A and Hudson, S R and Pfefferl{\'e}, D and
               Helander, P",
  journal   = "J. Plasma Phys.",
  publisher = "Cambridge University Press (CUP)",
  volume    =  87,
  number    =  2,
  month     =  apr,
  year      =  2021,
  language  = "en"
}

@ARTICLE{intro_single_nae,
  title     = "Single-stage gradient-based stellarator coil design:
               Optimization for near-axis quasi-symmetry",
  author    = "Giuliani, Andrew and Wechsung, Florian and Cerfon, Antoine and
               Stadler, Georg and Landreman, Matt",
  journal   = "J. Comput. Phys.",
  publisher = "Elsevier BV",
  volume    =  459,
  number    =  111147,
  pages     = "111147",
  month     =  jun,
  year      =  2022,
  language  = "en"
}

@ARTICLE{formulation_IXB,
  author={Strickland, N. M. and Wimbush, S. C.},
  journal={IEEE Transactions on Applied Superconductivity}, 
  title={The Magnetic-Field Dependence of Critical Current: What We Really Need to Know}, 
  year={2017},
  volume={27},
  number={4},
  pages={1-5},
  doi={10.1109/TASC.2016.2636561}}

@ARTICLE{formulation_IXB2,
  title     = "High field Ic characterizations of commercial {HTS} conductors",
  author    = "Miyoshi, Y and Nishijima, G and Kitaguchi, H and Chaud, X",
  journal   = "Physica C Supercond.",
  publisher = "Elsevier BV",
  volume    =  516,
  pages     = "31--35",
  month     =  sep,
  year      =  2015
}

@article{landreman_simsopt:_2021,
	title        = {SIMSOPT: A flexible framework for stellarator optimization},
	author       = {Landreman, Matt and Medasani, Bharat and Wechsung, Florian and Giuliani, Andrew and Jorge, Rogerio and Zhu, Caoxiang},
	year         = 2021,
	month        = sep,
	journal      = {Journal of Open Source Software},
	volume       = 6,
	number       = 65,
	pages        = 3525,
	doi          = {10.21105/joss.03525},
	issn         = {2475-9066},
	url          = {https://joss.theoj.org/papers/10.21105/joss.03525},
	urldate      = {2024-05-21}
}

@article{conlin2024stellarator,
  title={Stellarator optimization with constraints},
  author={Conlin, Rory and Kim, Patrick and Dudt, Daniel W and Panici, Dario and Kolemen, Egemen},
  journal={Journal of Plasma Physics},
  volume={90},
  number={5},
  pages={905900501},
  year={2024},
  publisher={Cambridge University Press}
}

@article{wechsung2022precise,
  title={Precise stellarator quasi-symmetry can be achieved with electromagnetic coils},
  author={Wechsung, Florian and Landreman, Matt and Giuliani, Andrew and Cerfon, Antoine and Stadler, Georg},
  journal={Proceedings of the National Academy of Sciences},
  volume={119},
  number={13},
  pages={e2202084119},
  year={2022},
  publisher={National Academy of Sciences}
}

@article{kaptanoglu2025reactor,
  title={Reactor-scale stellarators with force and torque minimized dipole coils},
  author={Kaptanoglu, Alan A and Wiedman, Alexander and Halpern, Jacob and Hurwitz, Siena and Paul, Elizabeth J and Landreman, Matt},
  journal={Nuclear Fusion},
  volume={65},
  number={4},
  pages={046029},
  year={2025},
  publisher={IOP Publishing}
}

@article{hurwitz2025electromagnetic,
  title={{Electromagnetic coil optimization for reduced Lorentz forces}},
  author={Hurwitz, Siena and Landreman, Matt and Huslage, Paul and Kaptanoglu, Alan},
  journal={Nuclear Fusion},
  volume={65},
  number={5},
  pages={056044},
  year={2025},
  publisher={IOP Publishing}
}

@article{jang2025exponential,
  title={Exponential Spectral Scaling: Robust and Efficient Stellarator Boundary Optimization via Mode-Dependent Scaling},
  author={Jang, Byoungchan and Conlin, Rory and Landreman, Matt},
  journal={arXiv preprint arXiv:2509.16320},
  year={2025}
}

@ARTICLE{aries_cs,
  title     = "Physics design for {ARIES-CS}",
  author    = "Ku, L P and Garabedian, P R and Lyon, J and Turnbull, A and
               Grossman, A and Mau, T K and Zarnstorff, M and {ARIES Team}",
  journal   = "Fusion Sci. Technol.",
  publisher = "Informa UK Limited",
  volume    =  54,
  number    =  3,
  pages     = "673--693",
  month     =  oct,
  year      =  2008,
  language  = "en"
}

@article{gil2025augmented,
  title={{Augmented Lagrangian methods produce cutting-edge magnetic coils for stellarator fusion reactors}},
  author={Gil, Pedro F and Kaptanoglu, Alan A and Stenson, Eve V},
  journal={arXiv preprint arXiv:2507.12681},
  year={2025}
}

@article{zhu2017new,
  title={New method to design stellarator coils without the winding surface},
  author={Zhu, Caoxiang and Hudson, Stuart R and Song, Yuntao and Wan, Yuanxi},
  journal={Nuclear Fusion},
  volume={58},
  number={1},
  pages={016008},
  year={2017},
  publisher={IOP Publishing}
}

@article{robin2022minimization,
  title={Minimization of magnetic forces on stellarator coils},
  author={Robin, R{\'e}mi and Volpe, Francesco A},
  journal={Nuclear Fusion},
  volume={62},
  number={8},
  pages={086041},
  year={2022},
  publisher={IOP Publishing}
}

@article{kappel_magnetic_2024,
	title        = {The magnetic gradient scale length explains why certain plasmas require close external magnetic coils},
	author       = {Kappel, John Thomas and Landreman, M and Malholtra, Dhairya},
	year         = 2024,
	month        = jan,
	journal      = {Plasma Physics and Controlled Fusion},
	doi          = {10.1088/1361-6587/ad1a3e},
	issn         = {0741-3335, 1361-6587},
	url          = {https://iopscience.iop.org/article/10.1088/1361-6587/ad1a3e},
	urldate      = {2024-01-08},
}

@article{drevlak_automated_1998,
	title        = {Automated Optimization of Stellarator Coils},
	author       = {Drevlak, Michael},
	year         = 1998,
	month        = mar,
	journal      = {Fusion Technology},
	volume       = 33,
	number       = 2,
	pages        = {106--117},
	doi          = {10.13182/FST98-A21},
	issn         = {0748-1896},
	url          = {https://www.tandfonline.com/doi/full/10.13182/FST98-A21},
	urldate      = {2023-11-03},
}

@article{boozer_optimization_2000,
	title        = {Optimization of the current potential for stellarator coils},
	author       = {Boozer, Allen H.},
	year         = 2000,
	month        = feb,
	journal      = {Physics of Plasmas},
	volume       = 7,
	number       = 2,
	pages        = {629--634},
	doi          = {10.1063/1.873849},
	issn         = {1070-664X, 1089-7674},
	url          = {https://pubs.aip.org/pop/article/7/2/629/457766/Optimization-of-the-current-potential-for},
	urldate      = {2023-08-28},
}

@article{najmabadi2008aries,
  title={{The ARIES-CS compact stellarator fusion power plant}},
  author={Najmabadi, F and Raffray, AR and Abdel-Khalik, SI and Bromberg, L and Crosatti, L and El-Guebaly, L and Garabedian, PR and Grossman, AA and Henderson, D and Ibrahim, A and others},
  journal={Fusion Science and Technology},
  volume={54},
  number={3},
  pages={655--672},
  year={2008},
  publisher={Taylor \& Francis}
}

@article{qian2022simpler,
  title={Simpler optimized stellarators using permanent magnets},
  author={Qian, T and Zarnstorff, M and Bishop, D and Chamblis, A and Dominguez, A and Pagano, C and Patch, D and Zhu, C},
  journal={Nuclear Fusion},
  volume={62},
  number={8},
  pages={084001},
  year={2022},
  publisher={IOP Publishing}
}

@article{qian2023design,
  title={{Design and construction of the MUSE permanent magnet stellarator}},
  author={Qian, TM and Chu, X and Pagano, C and Patch, D and Zarnstorff, MC and Berlinger, B and Bishop, D and Chambliss, A and Haque, M and Seidita, D and others},
  journal={Journal of Plasma Physics},
  volume={89},
  number={5},
  pages={955890502},
  year={2023},
  publisher={Cambridge University Press}
}

@article{landreman2021stellarator,
  title={Stellarator optimization for good magnetic surfaces at the same time as quasisymmetry},
  author={Landreman, Matt and Medasani, Bharat and Zhu, Caoxiang},
  journal={Physics of Plasmas},
  volume={28},
  number={9},
  year={2021},
  publisher={AIP Publishing}
}

@article{pomphrey_innovations_2001,
	title = {Innovations in compact stellarator coil design},
	volume = {41},
	issn = {0029-5515},
	url = {https://doi.org/10.1088/0029-5515/41/3/312},
	doi = {10.1088/0029-5515/41/3/312},
	language = {en},
	number = {3},
	urldate = {2026-02-07},
	journal = {Nuclear Fusion},
	author = {Pomphrey, N. and Berry, L. and Boozer, A. and Brooks, A. and Hatcher, R. E. and Hirshman, S. P. and Ku, L.-P. and Miner, W. H. and Mynick, H. E. and Reiersen, W. and Strickler, D. J. and Valanju, P. M.},
	month = mar,
	year = {2001},
	pages = {339},
}

@article{landreman2017improved,
  title={An improved current potential method for fast computation of stellarator coil shapes},
  author={Landreman, Matt},
  journal={Nuclear Fusion},
  volume={57},
  number={4},
  pages={046003},
  year={2017},
  publisher={IOP Publishing}
}

@article{merkel1987solution,
  title={Solution of stellarator boundary value problems with external currents},
  author={Merkel, Peter},
  journal={Nuclear Fusion},
  volume={27},
  number={5},
  pages={867},
  year={1987},
  publisher={IOP Publishing}
}

@article{shor_quadratic_1987,
	title        = {Quadratic optimization problems},
	author       = {Shor, Naum Z},
	year         = 1987,
	journal      = {Soviet Journal of Computer and Systems Sciences},
	volume       = 25,
	pages        = {1--11}
}

@inproceedings{strykowsky_engineering_2009,
	title        = {Engineering cost schedule lessons learned on NCSX},
	author       = {Strykowsky, R.L. and Brown, T. and Chrzanowski, J. and Cole, M. and Heitzenroeder, P. and Neilson, G.H. and Rej, Donald and Viol, M.},
	year         = 2009,
	month        = jun,
	booktitle    = {2009 23rd {IEEE}/{NPSS} {Symposium} on {Fusion} {Engineering}},
	publisher    = {IEEE},
	address      = {San Diego, CA, USA},
	pages        = {1--4},
	doi          = {10.1109/FUSION.2009.5226449},
	isbn         = {978-1-4244-2635-5},
	url          = {http://ieeexplore.ieee.org/document/5226449/},
	urldate      = {2024-05-01},
}

@phdthesis{elder_three-dimensional_2024,
	title        = {Three-dimensional magnetic fields: from coils to reconnection},
	author       = {Elder, Todd M.},
	year         = 2024,
	doi          = {10.7916/hket-nf64},
	url          = {https://doi.org/10.7916/hket-nf64},
	urldate      = {2024-05-21},
	school       = {Columbia University}
}

@article{dudt2023desc,
  title={{The DESC stellarator code suite Part 3: Quasi-symmetry optimization}},
  author={Dudt, Daniel W and Conlin, Rory and Panici, Dario and Kolemen, Egemen},
  journal={Journal of Plasma Physics},
  volume={89},
  number={2},
  pages={955890201},
  year={2023},
  publisher={Cambridge University Press}
}

@ARTICLE{VMEC_free_86,
  title     = "Three-dimensional free boundary calculations using a spectral
               Green's function method",
  author    = "Hirshman, S P and van RIJ, W I and Merkel, P",
  journal   = "Comput. Phys. Commun.",
  publisher = "Elsevier BV",
  volume    =  43,
  number    =  1,
  pages     = "143--155",
  month     =  dec,
  year      =  1986,
  language  = "en"
}

@article{conlin2024high,
  title={{High order free boundary MHD equilibria in DESC}},
  author={Conlin, Rory and Schilling, Jonathan and Dudt, Daniel W and Panici, Dario and Jorge, Rogerio and Kolemen, Egemen},
  journal={arXiv preprint arXiv:2412.05680},
  year={2024}
}

@article{dudt_desc_2020,
	title        = {DESC: A stellarator equilibrium solver},
	author       = {Dudt, D. W. and Kolemen, E.},
	year         = 2020,
	month        = oct,
	journal      = {Physics of Plasmas},
	volume       = 27,
	number       = 10,
	pages        = 102513,
	doi          = {10.1063/5.0020743},
	issn         = {1070-664X, 1089-7674},
	url          = {http://aip.scitation.org/doi/10.1063/5.0020743},
	urldate      = {2023-04-09}
}

@manual{mosek,
	title        = {The MOSEK optimization toolbox for MATLAB manual. Version 10.1.},
	author       = {MOSEK ApS},
	year         = 2024,
	url          = {http://docs.mosek.com/latest/toolbox/index.html}
}

@article{Henneberg_single_2021,
	title        = {Combined plasma–coil optimization algorithms},
	author       = {Henneberg, S. A. and Hudson, S. R. and Pfefferlé, D. and Helander, P.},
	year         = 2021,
	journal      = {Journal of Plasma Physics},
	volume       = 87,
	number       = 2,
	pages        = 905870226,
	doi          = {10.1017/S0022377821000271}
}

@software{jax2018github,
  author = {James Bradbury and Roy Frostig and Peter Hawkins and Matthew James Johnson and Chris Leary and Dougal Maclaurin and George Necula and Adam Paszke and Jake Vander{P}las and Skye Wanderman-{M}ilne and Qiao Zhang},
  title = {{JAX}: composable transformations of {P}ython+{N}um{P}y programs},
  url = {http://github.com/google/jax},
  version = {0.3.13},
  year = {2018},
}

@article{smiet,
author = {Smiet, C. B. and Loizu, J. and Balkovic, E. and Baillod, A.},
title = {Efficient single-stage optimization of islands in finite-beta stellarator equilibria},
journal = {Physics of Plasmas},
volume = {32},
number = {1},
pages = {012504},
year = {2025},
month = {01},
issn = {1070-664X},
doi = {10.1063/5.0226402},
url = {https://doi.org/10.1063/5.0226402},
eprint = {https://pubs.aip.org/aip/pop/article-pdf/doi/10.1063/5.0226402/20356029/012504_1_5.0226402.pdf},
}

@MISC{Lanke2025-re,
  title     = "{QUADCOIL} Quasi-single-stage Dataset",
  author    = "Lanke, Fu",
  year         = 2026,
  month        = April,
  publisher    = {Zenodo},
  doi          = {10.5281/zenodo.19923438},
  url          = {https://zenodo.org/records/19923438},
}

@ARTICLE{Henneberg2021-spectral_condensation,
  title     = "Representing the boundary of stellarator plasmas",
  author    = "Henneberg, S A and Helander, P and Drevlak, M",
  journal   = "J. Plasma Phys.",
  publisher = "Cambridge University Press (CUP)",
  volume    =  87,
  number    =  5,
  month     =  oct,
  year      =  2021,
  language  = "en"
}

@ARTICLE{qss_carlton_regcoil,
  title     = "Computing the shape gradient of stellarator coil complexity with
               respect to the plasma boundary",
  author    = "Carlton-Jones, Arthur and Paul, Elizabeth J and Dorland, William",
  journal   = "J. Plasma Phys.",
  publisher = "Cambridge University Press (CUP)",
  volume    =  87,
  number    =  2,
  month     =  apr,
  year      =  2021,
  language  = "en"
}

@ARTICLE{qss_Yu2024-sr,
  title     = "Quasi-single-stage optimization for permanent magnet
               stellarators",
  author    = "Yu, Guodong and Liu, Ke and Qian, Tianyi and Xie, Yidong and
               Nie, Xianyi and Zhu, Caoxiang",
  journal   = "Nucl. Fusion",
  doi       = "10.1088/1741-4326/ad521c",
  publisher = "IOP Publishing",
  volume    =  64,
  number    =  7,
  pages     = "076055",
  month     =  jul,
  year      =  2024,
  copyright = "http://creativecommons.org/licenses/by/4.0"
}

@misc{qss_barratt_differentiability_2019,
	title = {On the {Differentiability} of the {Solution} to {Convex} {Optimization} {Problems}},
	url = {http://arxiv.org/abs/1804.05098},
	doi = {10.48550/arXiv.1804.05098},
	language = {en},
	urldate = {2025-04-25},
	publisher = {arXiv},
	author = {Barratt, Shane},
	month = nov,
	year = {2019},
	note = {arXiv:1804.05098 [math]},
}

@inproceedings{agrawal_differentiable_2019,
	title = {Differentiable {Convex} {Optimization} {Layers}},
	volume = {32},
	url = {https://proceedings.neurips.cc/paper/2019/hash/9ce3c52fc54362e22053399d3181c638-Abstract.html},
	urldate = {2026-02-07},
	booktitle = {Advances in {Neural} {Information} {Processing} {Systems}},
	publisher = {Curran Associates, Inc.},
	author = {Agrawal, Akshay and Amos, Brandon and Barratt, Shane and Boyd, Stephen and Diamond, Steven and Kolter, J. Zico},
	year = {2019},
}

@article{pan_bpqp_2024,
	title = {{BPQP}: {A} {Differentiable} {Convex} {Optimization} {Framework} for {Efficient} {End}-to-{End} {Learning}},
	volume = {37},
	shorttitle = {{BPQP}},
	url = {https://proceedings.neurips.cc/paper_files/paper/2024/hash/8db12f7214d3a1a0c450ba751163e0fd-Abstract-Conference.html},
	doi = {10.52202/079017-2463},
	language = {en},
	urldate = {2026-02-07},
	journal = {Advances in Neural Information Processing Systems},
	author = {Pan, Jianming and Ye, Zeqi and Yang, Xiao and Yang, Xu and Liu, Weiqing and Wang, Lewen and Bian, Jiang},
	month = dec,
	year = {2024},
	pages = {77468--77493},
}

@inproceedings{lorraine_optimizing_2020,
	title = {Optimizing {Millions} of {Hyperparameters} by {Implicit} {Differentiation}},
	issn = {2640-3498},
	url = {https://proceedings.mlr.press/v108/lorraine20a.html},
	language = {en},
	urldate = {2026-02-07},
	booktitle = {Proceedings of the {Twenty} {Third} {International} {Conference} on {Artificial} {Intelligence} and {Statistics}},
	publisher = {PMLR},
	author = {Lorraine, Jonathan and Vicol, Paul and Duvenaud, David},
	month = jun,
	year = {2020},
	pages = {1540--1552},
}

@phdthesis{qss_paul_adjoint_nodate,
author={Paul,Elizabeth J.},
year={2020},
school={University of Maryland, College Park},
title={Adjoint Methods for Stellarator Shape Optimization and Sensitivity Analysis},
journal={ProQuest Dissertations and Theses},
pages={292},
isbn={9798678113061},
language={English},
url={https://www.proquest.com/dissertations-theses/adjoint-methods-stellarator-shape-optimization/docview/2454631007/se-2},
}

@article{drevlak_coil_1998,
	title = {Coil designs for a quasi-axially symmetric stellarator},
	url = {https://www.osti.gov/etdeweb/biblio/20045460},
	language = {English},
	urldate = {2026-02-07},
	author = {Drevlak, M.},
	month = jul,
	year = {1998},
}

@book{imbert-gerard_introduction_2024,
	address = {Philadelphia, PA},
	title = {An {Introduction} to {Stellarators}: {From} {Magnetic} {Fields} to {Symmetries} and {Optimization}},
	url = {https://epubs.siam.org/doi/abs/10.1137/1.9781611978223},
	doi = {10.1137/1.9781611978223},
	publisher = {Society for Industrial and Applied Mathematics},
	author = {Imbert-Gérard, Lise-Marie and Paul, Elizabeth J. and Wright, Adelle M.},
	year = {2024},
	note = {\_eprint: https://epubs.siam.org/doi/pdf/10.1137/1.9781611978223},
}

@article{kaptanoglu_greedy_2023,
	title = {Greedy permanent magnet optimization},
	volume = {63},
	issn = {0029-5515},
	url = {https://doi.org/10.1088/1741-4326/acb4a9},
	doi = {10.1088/1741-4326/acb4a9},
	language = {en},
	number = {3},
	urldate = {2026-02-07},
	journal = {Nuclear Fusion},
	publisher = {IOP Publishing},
	author = {Kaptanoglu, Alan A. and Conlin, Rory and Landreman, Matt},
	month = feb,
	year = {2023},
	pages = {036016},
}

@book{nesterov_interior-point_1994,
	title = {Interior-{Point} {Polynomial} {Algorithms} in {Convex} {Programming}},
	isbn = {978-0-89871-319-0 978-1-61197-079-1},
	url = {http://epubs.siam.org/doi/book/10.1137/1.9781611970791},
	doi = {10.1137/1.9781611970791},
	language = {en},
	urldate = {2026-02-07},
	publisher = {Society for Industrial and Applied Mathematics},
	author = {Nesterov, Yurii and Nemirovskii, Arkadii},
	month = jan,
	year = {1994},
	doi = {10.1137/1.9781611970791},
}

\end{document}